\PassOptionsToPackage{table,xcdraw}{xcolor}
\documentclass[sigconf]{acmart}

\setcopyright{none}
\AtBeginDocument{%
  \providecommand\BibTeX{{%
    \normalfont B\kern-0.5em{\scshape i\kern-0.25em b}\kern-0.8em\TeX}}}

\copyrightyear{}
\acmYear{}
\acmDOI{}

\acmConference[]{}{}{}
\acmBooktitle{}
\acmPrice{}
\acmISBN{}

\acmSubmissionID{6166}

\newcommand{\EXMA}{\textit{XEM}$^{2}$}

\usepackage{multirow}

\begin{document}

%%
%% The "title" command has an optional parameter,
%% allowing the author to define a "short title" to be used in page headers.
\title[{XEM}$^{2}$: Exercise Exertion Muscle-work Monitor]{Put Your Muscle Into It: Introducing {XEM}$^{2}$, a Novel Approach for Monitoring Exertion in Stationary Physical Exercises Leveraging Muscle Work}

%%
%% The "author" command and its associated commands are used to define
%% the authors and their affiliations.
%% Of note is the shared affiliation of the first two authors, and the
%% "authornote" and "authornotemark" commands
%% used to denote shared contribution to the research.
\author{Jana Franceska Funke}
\email{jana.funke@uni-ulm.de}
\orcid{0000-0002-0635-5078}
\affiliation{%
  \institution{Institute of Media Informatics, Ulm University}
  \city{Ulm}
  \country{Germany}
}

\author{Mario Sagawa}
\email{mario-1.sagawa@uni-ulm.de}
\orcid{0009-0008-2754-1487}
\affiliation{%
  \institution{Institute of Media Informatics, Ulm University}
  \city{Ulm}
  \country{Germany}
}

\author{Georgious Nurcan-Georgiou}
\email{georgios.nurcan-georgiou@uni-ulm.de}
\orcid{0009-0005-5503-8967}
\affiliation{%
  \institution{Institute of Media Informatics, Ulm University}
  \city{Ulm}
  \country{Germany}
  }

\author{Naomi Sagawa}
\email{naomi-1.sagawa@uni-ulm.de}
\orcid{0009-0005-1206-2334}
\affiliation{%
  \institution{Institute of Media Informatics, Ulm University}
  \city{Ulm}
  \country{Germany}
  }

\author{Dennis Dietz}
\orcid{0000-0002-4428-6544}
\affiliation{%
  \institution{LMU Munich}
  \city{Munich}
  \country{Germany}}
\email{dennis.dietz@ifi.lmu.de}

\author{Evgeny Stemasov}
\orcid{0000-0002-3748-6441}
\email{evgeny.stemasov@uni-ulm.de}
\affiliation{%
  \institution{Institute of Media Informatics, Ulm University}
  \city{Ulm}
  \country{Germany}
  }

\author{Enrico Rukzio}
\email{enrico.rukzio@uni-ulm.de}
\orcid{0000-0002-4213-2226}
\affiliation{%
  \institution{Institute of Media Informatics, Ulm University}
  \city{Ulm}
  \country{Germany}
}

\author{Teresa Hirzle}
\orcid{0000-0002-7909-7639}
\affiliation{%
  \institution{Department of Computer Science, University of Copenhagen}
  \city{Copenhagen}
  \country{Denmark}}
\email{tehi@di.ku.dk}

\renewcommand{\shortauthors}{Funke et al.}

%%
%% The code below is generated by the tool at http://dl.acm.org/ccs.cfm.
%% Please copy and paste the code instead of the example below.
%%
\begin{CCSXML}
<ccs2012>
   <concept>
       <concept_id>10003120.10003121.10011748</concept_id>
       <concept_desc>Human-centered computing~Human computer interaction (HCI)</concept_desc>
       <concept_significance>500</concept_significance>
       </concept>
 </ccs2012>
 <concept>
<concept_id>10003120.10003121.10003129</concept_id>
<concept_desc>Human-centered computing~Interactive systems and tools</concept_desc>
<concept_significance>500</concept_significance>
</concept>
\end{CCSXML}

\ccsdesc[500]{Human-centered computing}
\ccsdesc[500]{Human-centered computing~Human computer interaction (HCI)}
\ccsdesc[300]{Human-centered computing~Interactive systems and tools}

%%
%% Keywords. The author(s) should pick words that accurately describe
%% the work being presented. Separate the keywords with commas.
\keywords{physical exercise, body tracking, muscle work, muscle engagement, movement measure, burned calories}
%%
%% The abstract is a short summary of the work to be presented in the
%% article.
\begin{abstract}

We present a novel system for camera-based measurement and visualization of muscle work based on the Hill-Type-Muscle-Model: the exercise exertion muscle-work monitor (\textit{XEM}$^{2}$).
Our aim is to complement and, thus, address issues of established measurement techniques that offer imprecise data for non-uniform movements (burned calories) or provide limited information on strain across different body parts (self-perception scales).
We validate the reliability of  XEM's measurements through a technical evaluation of ten participants and five exercises.
Further, we assess the acceptance, usefulness, benefits, and opportunities of  \textit{XEM}$^{2}$ in an empirical user study.
Our results show that  \textit{XEM}$^{2}$ provides reliable values of muscle work and supports participants in understanding their workout while also providing reliable information about perceived exertion per muscle group.
With this paper, we introduce a novel system capable of measuring and visualizing exertion for single muscle groups, which has the potential to improve exercise monitoring to prevent unbalanced workouts.

\end{abstract}

%% A "teaser" image appears between the author and affiliation
%% information and the body of the document, and typically spans the
%% page.
\begin{teaserfigure}
  \includegraphics[width=\textwidth]{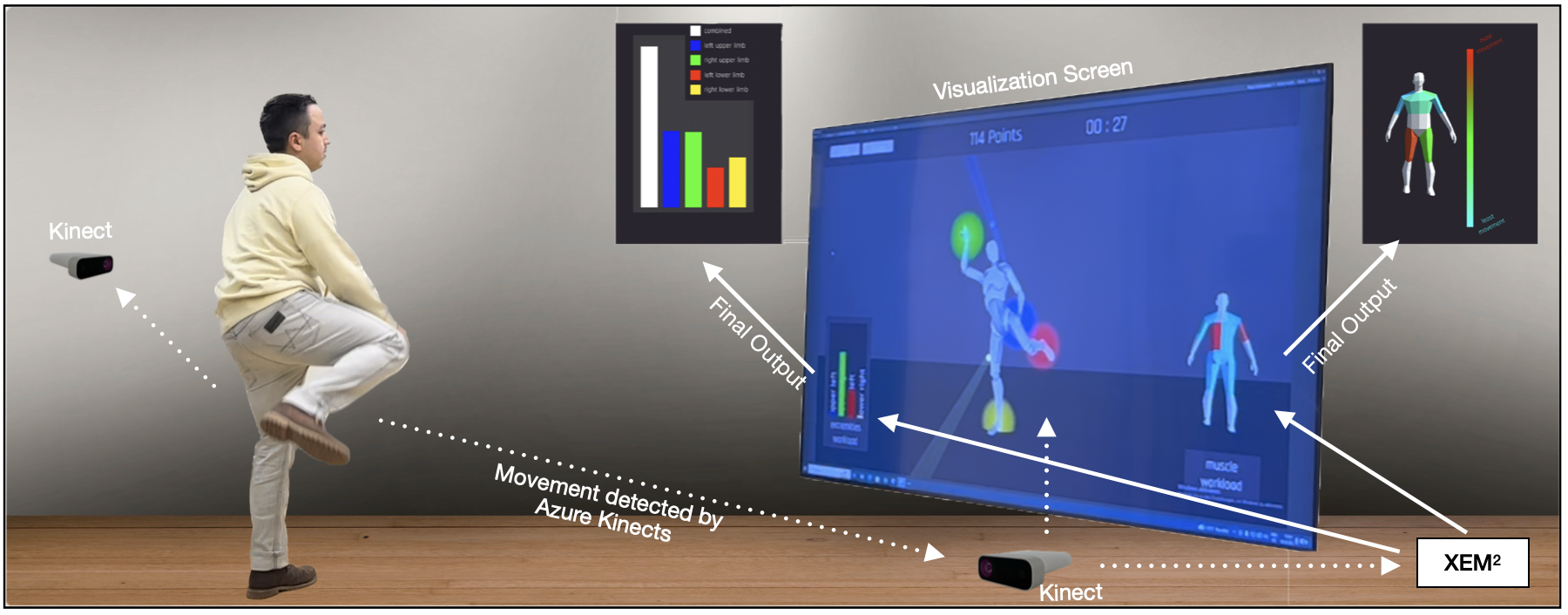}
  \caption{The \EXMA setup: On the left, a participant is playing an exercise game (exergame). Their movements are detected by two Azure Kinect Cameras. We then use our \EXMA~ system to evaluate the exertion that different muscle groups need to perform that movement. The results are visualized in two different ways, a bar chart for prime movers (right/left upper/lower limb) and a 3D avatar with highlighted muscle groups (M. Deltoideus, M. Pectoralis Major, M. Triceps Brachii, M. Biceps Brachii and M. Brachialis, M. Latissimus Dorsi, M. Gluteus Maximus, Mm. ischiocrurales, M. Quadriceps Femoris).}
  \Description{The picture shows a person doing exercises in front of a large screen. On the side and underneath the screen is an Azure Kinect positioned. between the person and the Kinects are dotted lines saying that the movement is detected by the Azure Kinects. From the Azure Kinect, there is a dotted line towards a box saying XEM and towards the middle avatar figure on the screen. From the XEM box there are two arrows toward the left and right side of the screen, where we can see the limb exertion boxplot and the muscle group exertion highlighted avatar. Both create a final visualization shown with a labelled arrow saying final output. The final output is the accumulated value over the whole session.}
  \label{fig:teaser}
\end{teaserfigure}

% \received{20 February 2007}
% \received[revised]{12 March 2009}
% \received[accepted]{5 June 2009}

%%
%% This command processes the author and affiliation and title
%% information and builds the first part of the formatted document.
\maketitle

\section{Introduction} 
Monitoring physical exercise can help people achieve their fitness goals by fostering motivation.
Tracking activity over time helps people set goals, allowing them to stay accountable in their fitness journey \cite{mopasTrainingFeelWearable2020,vargemidisWearablePhysicalActivity2020,carrieri-kohlmanExerciseTrainingDecreases1996}.
Besides supporting a person in following a healthy lifestyle, physical exercise monitoring permits early interventions in some cases of physical impairment (e.g., injuries), which can reduce high costs in the health sector and alleviate adverse long-term effects.
Personal monitoring is especially important since physical exercise often no longer takes place in a supervised environment but increasingly alone and at home \cite{thebusinessresearchcompanyOnlineVirtualFitness2023}. 

% Solution: What is the solution that we came up with to solve it?
%In this paper, we present a novel approach for physical exercise monitoring that - in contrast to many established measurement methods like burned calories - provides detailed measurements of exertion in single muscle groups rather than a generic measure.
%The \textit{Exercise Exertion Muscle-work Monitor} (\EXMA) is based on a camera-based movement tracking system, and measures and visualizes muscle work of different muscle groups and their variation over time.
% In contrast
In this work, we present a novel system for exertion measurement and visualization: the \textit{Exercise Exertion Muscle-work Monitor} (\EXMA). \EXMA is a camera-based movement tracking system that measures and visualizes muscle work of different muscle groups and their variation over time. 
The system is based on a musculoskeletal calculation using the Hill-Type Muscle Model~\cite{zhouRoughPreciseHumanInspired2019} and the Kinesis Unity Plugin for calculating exertion. 
We use the term \textit{muscle work} since we focus on the work a muscle has to accomplish to move or hold a position, though it is not the exact physical definition of work. 
\EXMA~ provides a visualization of the used muscles both during and after an exercise has been performed. 
Since \EXMA~ measures exertion on a muscle-specific level, it offers advantages over other exertion measurement techniques by providing detailed information. 
Nevertheless, we emphasize that \EXMA~ was not designed to replace other measures but rather to complement them.

One established way to measure physical exercise is burned calories~\cite{ainsworthCompendiumPhysicalActivities2011}.
Burned calories are calculated based on a single generic value called the \textit{metabolic equivalent (MET)}, a person's body parameters (age, weight, heart rate), and training time \cite{ainsworthCompendiumPhysicalActivities2000,be2011CompendiumPhysical2011,ainsworthCompendiumPhysicalActivities2011}.
The MET is medically pre-measured in detail for each exercise type based on oxygen uptake, measured in ml/kg/min.
This value is relatively accurate for representing repetitive and steady activities, where one can assume an even burn of calories, such as running, walking, and swimming. 
However, it is rather imprecise for exercises like yoga or dancing where movements and intensity change frequently during the exercise, but only a single approximate value is applied \cite{ainsworthCompendiumPhysicalActivities2000}. %,
For exercise games (exergames), an emerging approach to home exercise motivation,  there are approaches using MET values \cite{vrheathinstituteMethodology2017}.
However, the associated MET value has to be pre-measured for each new game or game sequence. 
Secondly, there are self-perception scales like the Rating of Perceived Exertion scale (RPE) \cite{shariatBorgCR10Scale2018,williamsBorgRatingPerceived2017}. 
While they provide more detailed values, they are not as accurate as objective measures, because they need practice and good self-assessment \cite{hermanValidityReliabilitySession2006}.
Finally, burned calories and self-perception scales only provide a single value for the exertion of an exercise. A single value gives little information about the physical strain of specific body parts and about the muscles being involved. %If users would know about muscles involved they could balance activities and workouts to prevent muscle overload and prevent long-term consequences.
However, more information and support for physical exercise is often desired \cite{vargemidisWearablePhysicalActivity2020,maradaniRoleVisualizationTelerehabilitation2017}.

Our system, \EXMA, complements these generic exercise measurement methods by providing reliable information about individual muscles.
Thus, it can not only help users to have a well-balanced selection of exercises to avoid unilateral training but also prevent overload and injuries \cite{reedPracticalApproachesPrescribing2016,kiblerMusculoskeletalAdaptationsInjuries1992,dinubileStrengthTraining1991}.

We measured validity and evaluated the acceptance and usefulness of \EXMA~ in an empirical user study (N=36). 
First, we measured \EXMA~ values for five different exercises and compared the change of values for different muscle groups.
Secondly, we quantitatively compared participants' acceptance of \EXMA~ with two established exertion measurement methods: (1) the RPE scale and (2) burned calories. 
We also collected qualitative feedback from the participants about the advantages, disadvantages, and potential improvements of \EXMA.
Thirdly, we focused on the visualization of \EXMA~ and qualitatively evaluated the effects and benefits of the muscle work 
visualized by a 3D virtual avatar and box plot.

%5. Findings & Contribution 
%The findings of the first part of the user study demonstrate that \EXMA~ was about equal in providing useful, precise and trustworthy information about exertion compared to burned calories and the RPE scale. 
For technical validity, repetitive values showed only small fluctuations between repetitive movements but big differences between involving a muscle group or not.
The findings of the user study part demonstrate that in providing useful, precise, and trustworthy information about exertion \EXMA~ was about equal compared to burned calories and the RPE scale.
%The participants reported a high level of perceived trust, precision, and helpfulness for all three exercise measurement techniques ((1) RPE scale, (2) burned calories , (3) \EXMA).
%Although calories burned is a well-known concept and was rated very positively, many participants lack understanding of how it works. 
The participants found that \EXMA~ presented the most interesting and useful information about their workout. %was presented by \EXMA. 
%The second study part reveals that the participants perceived our visualizations to be non-distracting and useful for understanding the body during an exercise. 
%They also indicated that \EXMA~ would help them in achieving a balanced workout.
%An additional measurement of squats was used to compare values for validity.
In summary, \EXMA~ provides granular information about muscle work in specific muscles during any stationary physical home workout and can be used in conjunction with traditional measures to provide a more complete picture of exercise performance. 
However, we would like to note that the system is not medically approved, and the measurement exhibits some weaknesses, so that it is more suitable for private use as a monitor for a balanced exercise and additional measures in, for example, exercise studies.
In the future, we plan to further broaden the technical aspects of our system and envision transferring the system to smartphone use. 

%* burned calories is known but participants have mostly no idea how it works -> our tool no lack in perceived trust, precision, helpfulness 
%* most intersting and useful information has our tool 

%* \EXMA visualization not distracting, useful (auch statistisch) for understanding body in workout and balanced workout -> health , motivation, 

%* technisch schwer zu sagen ob genauer wie BC oder RPE aber soll ja auch nicht ersetzten, participants mögen ein komination und daher kann es einfach auch für granularere informationene genutzt werden 

%-> zukunft technisch verbesserung und userfull visualizations erweitern 

In summary, we contribute: 
\begin{itemize}
\item[(1)  ] A camera-based physical exercise muscle work monitoring system (\EXMA) that measures and visualizes the muscle work for different muscle groups with a 3D avatar and box plots for more precise information of a physical exercise;
\item[(2)  ] A technical evaluation to show the validity of the measurement.
\item[(3)  ] Insights from a user study with 36 participants: 
\begin{itemize}
    \item[(3a)] Part 1: Comparing the effects of different measurement techniques, we found that RPE, burned calories, and \EXMA~ are rated as approximately similarly helpful, trustworthy, and precise. However, \EXMA~ presented the most interesting information and understanding of the body 
    %\item[(3b)] Part 2: Our results indicate that \EXMA~ visualizations are useful for motivation, a balanced workout, and body understanding. Although \EXMA~ is not distracting for visualization while exercising, participants were not particularly excited about in-situational use. 
    \item[(3b)] A basic comparison between the measured values of RPE, burned calories, and \EXMA, highlighting their potential individual differences. 
\end{itemize}
\end{itemize}

\section{Related Work} 
Our work is mainly related to other measurement techniques for exertion in physical exercise and to visualizations supporting users' understanding of their physical tasks and preventing mistakes and injuries.

While a lot of technical, physical activity-related work focuses on tracking \cite{arakawaMIPoserHumanBody2023,chenEnvironmentawareMultipersonTracking2023} and activity detection \cite{saeedMultitaskSelfSupervisedLearning2019}, our work focuses on the activity measuring techniques itself and on how to use the measured output to create a benefit for users.
Specifically, we aim to provide meaningful feedback \emph{during} the activity while balancing detail (i.e., providing more than a single number) and simplicity.

\subsection{Measurement Techniques for Physical Exercise}

Physical movement is an essential component of a healthy lifestyle and crucial for preventing various chronic diseases, such as diabetes type 2, cardiovascular disease, or obesity \cite{leeEffectPhysicalInactivity2012}. 
Therefore, it is essential to monitor, for example, physical exercise in order to assess individuals' activity progress, push motivation through goal setting and competitive comparison to promote healthy behaviours.
One’s progress can be assessed through self-perception or technical measurements. 

One way to assess the physical exertion level is to ask about the effort exerted during a training session or physical exercise. The Rating of Perceived Exertion scale (RPE) provides a method used by athletes \cite{williamsBorgRatingPerceived2017}. On a scale of 6 to 20 (very, very light – very, very hard), the trainee indicates the perceived effort after a training session based on subjective perception. This scale's simplicity allows for a quick overview of the perceived effort. However, factors such as heart rate, gender, or respiration are not considered. 
A revised version of the RPE scale only ranges from 0 = "Rest" to 10 = "Maximal Effort" \cite{shariatBorgCR10Scale2018,vanderzwaardValidityReliabilityFacial2022}.
%While the RPE scale focuses on a single training session, other self-report questionnaires consider the total amount of physical activity performed over the past seven days (International Physical Activity Questionnaire (IPAQ)) or one “typical” week (Global Physical Activity Questionnaire (GPAQ)). Both IPAQ and GPAQ give metabolic equivalent (MET) scores for the work done \cite{ipaq2005, armstrong2006development}. 
Another well-established method to measure physical exercise intensity is the unit metabolic equivalent (MET) \cite{ainsworth2011compendium2011}. MET is defined as the amount of energy expended during physical activity relative to the resting metabolic rate, where 1 MET equals 3.5 ml O2/kg/min. Alternatively, MET can be defined by oxygen uptake in ml/kg/min. The Compendium of Physical Activities~\cite{ainsworthCompendiumPhysicalActivities2000}, a catalogue that provides energy expenditure values for various activities based on the MET unit, provides a standardized way to compare and assess physical activity across different populations and settings. However, the Compendium has limitations in the accuracy of its MET measurements since the values are one approximate value over time~\cite{KeadleErrors2010,ainsworthCompendiumPhysicalActivities2000}.  

Besides the method of self-assessment, technical measurements are another approach to quantify physical exertion (e.g., heart rate monitoring). The relationship between heart rate and energy expenditure is well-established, and various heart rate monitors are available for individuals to track their exertion~\cite{RennieEstimating2001}. 
%However, this method has limitations, such as its dependence on factors like age, gender, and medication use. Therefore, generalization is not possible.
A more common approach is combining these measures as burned calories~\cite{ainsworth2011compendium2011}. 
Burned Calories are calculated using the activity in minutes, the weight in kg, age, gender, heart rate, VO$_{2Max}$ and the MET value of the exercise. Some of the factors can be left out, which will result in a less accurate measure.

\EXMA~ is designed to complement these generic measures by providing detailed values for the exertion in specific muscles used during an exercise.

\subsection{Muscle-tracking Systems}

Overcoming the limitations of the Compendium and heart rate monitoring, alternative techniques have been proposed in the human-computer interaction (HCI) community to measure muscle contraction during physical activity. One such study by \citet{mokayaBurnoutWearableSystem2016} demonstrates the use of a surface electromyogram (sEMG), which measures the electrical activity generated by skeletal muscles during contraction. sEMG provides insight into the timing and activation level of different muscles during physical activity and can be used to monitor muscle fatigue and imbalances. %While sEMG are very receptive to noise produced by the muscle movement demonstrates incorporated accelerometers to counter these vibrations. 
The approach of \citet{mokayaBurnoutWearableSystem2016} relies on a network of accelerometers to record muscle vibrations while estimating body motion and using sEMG to acquire a ground truth measurement of muscle fatigue.

Another technique for measuring muscle contraction is the surface mechanomyogram (MMG), which measures the mechanical vibrations generated by muscle contractions. Compared to sEMG, MMG is less sensitive to external factors and can provide information on muscle force and fatigue \cite{WoodwardSegmenting2019}. Further, research suggests that a combination of both systems allows for improved signal quality \cite{hong-liuMMGSignalIts2010, mokayaMyoVibeEnablingInertial2018}.

The examples provided in this context often exhibit a high degree of specificity towards individual muscle groups, thereby disregarding the holistic nature of the human body. It is evident that the desired level of detail often hinges on questions of prioritization. This can be reflected in the visual presentation of information (e.g., visual hierarchy, omission of information). In the next section, we will discuss a set of systems that emphasize visualizing physical body metrics.

\subsection{Visualizations for Exercise and Movement}

Visualization can help users to understand complex and overwhelming data and derive goal-oriented actions. Therefore, health monitoring through digitization can bring many benefits, such as remote monitoring and long-term assessment \cite{pageVisualizationHealthMonitoring2015}. 
For instance, a work by \citet{boovaraghavanTAOContextDetection2023} uses their wellness prototype with contextual information paired with activity tracking to denote the stress and productivity of office occupants.
\citet{ryokaiCommunicatingInterpretingWearable2015} designed a system to visualize activity data (e.g., heart rate, steps, skin temperature) for interaction and communication with health coaches. It helped participants to understand their physical condition and activities better. 
Various approaches for supporting and visualizing exercise focus on situations where the trainer is absent or part-time absent. Nevertheless, the user has to progress with exercise without trainers (i.e., unsupervised). Here, supportive tools become indispensable. 
These tools can rely on conventional 2D displays or leverage immersive setups like VR (virtual reality) headsets. 
For example, MuscleRehab~\cite{zhuMuscleRehabImprovingUnsupervised2022} uses electrical impedance tomography (EIT) to visualize muscle engagement and a VR environment to monitor and improve the accuracy of therapeutic exercises.
The study found that with the muscle engagement visualization, users' accuracy of therapeutic exercise execution increased by 15\%~\cite{zhuMuscleRehabImprovingUnsupervised2022}.
\citet{kocielnikReflectionCompanionConversational2018} built an important tool called Reflection Companion for activity monitoring and reflection of fitness tracker data. Their Reflection Companion helped to increase motivation, empowerment, and adoption of new behaviours. Learning from these works, we assume that users are interested in monitoring their exercise, getting detailed and visually processed feedback, and may draw benefits from this feedback.

The current muscle-related methods lack granularity regarding the active muscle groups (e.g., burned calories), are often used ex-situ (e.g., Exercise Tracking Apps), or are designed for experts. (e.g.,~\cite{lamAutomatedRehabilitationSystem2016}). 

\EXMA~ uses two types of visualization (a 3D avatar highlighting different muscles and a box plot) to communicate the exertion of different muscle parts to users.

\section[Exercise Exertion Muscle-Work Monitor]{Exercise Exertion Muscle-Work Monitor (\EXMA)} 
The \textit{Exercise Exertion Muscle-Work Monitor} (\EXMA) is a camera-based exertion monitor system that measures and visualizes muscle work for individual muscle groups and their variation over time and during a stationary physical exercise. 
Our system was developed in consultation with a physiotherapist who is part of the author team. 
The physiotherapist was involved in all discussions. They were part of the development team and thus influenced the design process of \EXMA~ with domain knowledge about activity measurement techniques and their boundaries, the human body, and muscle functionality. This entailed understanding and discussing the problem, implementing the tool, and discussing the effects and limitations of \EXMA. However, there was no formal documentation of the physiotherapist's involvement, such as interviews or design guidelines.

\subsection{System Setup}
\EXMA~ consists of a camera-based motion capture system (two Azure Kinects), a visualization screen, and the muscle activity measurement software itself (see \autoref{fig:teaser}).

\subsubsection{Camera-based Motion Capture System}
We used two Microsoft Azure Kinects\footnote{\url{https://azure.microsoft.com/de-de/products/kinect-dk}} for the camera-based motion tracking. 
While one Azure Kinect Camera is sufficient, two are more precise. 
The Azure Kinect is equipped with a 1-megapixel depth sensor (wide and narrow field-of-view), 
a 12-megapixel RGB video camera, an accelerometer and gyroscope (IMU), and external sync pins to synchronize sensor streams from multiple Kinect devices. 
The depth camera for body tracking is operating in a Narrow Field of View Unbinned (NFOV) configuration with 60 frames per second (FPS), a resolution of 640x576, and an operating depth of about 0.5 - 3.86 m. 
One camera is relatively small (103 x 39 x 126 mm) and light (440 g). 
%results in a 3D precision as fine as Xmm at long ranges and Xmm in close proximity
We found that because the camera partly had problems recognizing the body with very dark clothes, relatively bright clothes should be worn for ideal tracking performance.
We use the \emph{Azure Kinect Body Tracking SDK} for recording, streaming, and detecting the skeleton data as an output using the depth sensor and RGB video camera. 
The motion data of the skeleton over time is processed with the cross-platform game engine Unity, where it is the basis of our muscle work calculations and visualizations.
%Simultaneously, the motion data was used for a self-implemented exergame in study part two where participants moved a 3D Avatar mirroring the participant's movements. 

\subsubsection{ \EXMA~ Visualization Screen}
The visualization screen is implemented with the Unity Game Engine (Version 2021.3.2f1) as a PC desktop application. 
For our setup, we used a 75-inch screen (\autoref{fig:teaser}) to show visualizations and supporting screens for study execution.
The screen content consists of three core parts. %containing mostly three parts.
First, the box plots on the right show the muscle movement for the four limbs. 

The left upper limb is shown in blue, the right upper limb in green, the left lower limb in yellow, and the right lower limb in red and white depicting overall combined value. 
The colours are not only applied to the bar charts but also to the end of the body limbs as hands and feet on the avatar in the middle in \autoref{fig:teaser}. 
The avatar in the middle is the representation of the participant. 
It mirrors the user and moves accordingly.
On the right, an abstracted 3-dimensional (3D) human model is shown, highlighting the body parts referring to the work of the user's muscle groups. The muscle groups we decided to visualize are depicted in \autoref{fig:kinesismusclemodelandMuscleGroups}. The human model is turning to show all perspectives of the muscle groups. 
The colour gradient ranges from turquoise (low muscle work) over green to red (high muscle work), depending on the relative amount of movement executed with that muscle. 
The arrangement of visualizations and additional information screens differ in the study. 
They were adjusted to fit the study setup and research question. 
We will present the details in the \autoref{lab:userstudy}.

\subsubsection{Visualizing Muscle Work}
\emph{XEM$^2$} uses the movement data recorded by the Azure Kinect to calculate the muscle work that had been necessary to execute the movement. 
The detected user movement with the camera is applied to the Kinesis virtual avatar model with a detailed muscle structure. 
Using the movement of the avatars' muscles, the muscle work for this muscle will be calculated. Thereupon, many muscles can be combined into one muscle group again. 
Depending on the value, a colour on the colour gradient will be chosen for the highlighted avatar model, and the value will likewise be applied to box plots.

%The amount of muscle work is visualized by colour-grading muscle groups of an avatar and showing box plots for limbs (\autoref{fig:kinesismusclemodelandMuscleGroups}). 

\begin{figure}[h]
    \centering
    \includegraphics[width=0.5\textwidth]{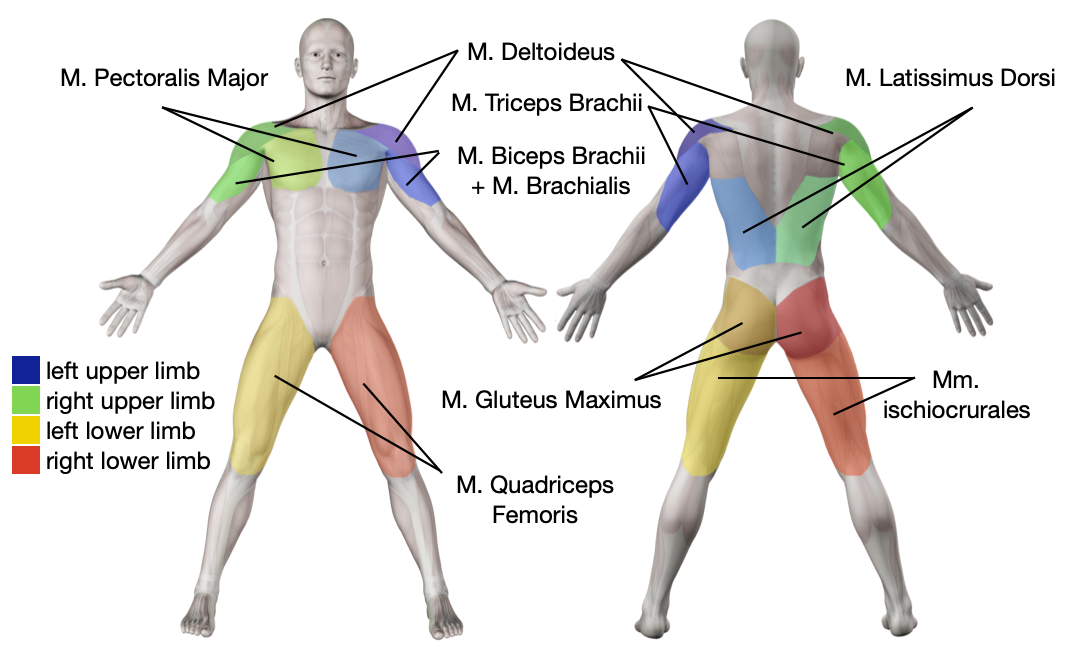} 
    \caption{The human model represents the muscle groups we decided to visualize as they are primarily involved in the movement of upper and lower limbs (i.e., prime movers). 
    The coloured areas on the human model are labelled with the muscle group name, while the colour indicates the affiliation to the four limbs.}
    \Description{The figure shows a muscle model of a person from the front and back. While all the muscle groups we created are labelled with their medical terms (M. Deltoideus, M. Pectoralis Major, M. Triceps Brachii, M. Biceps Brachii and M. Brachialis, M. Latissimus Dorsi, M. Gluteus Maximus, Mm. ischiocrurales, M. Quadriceps Femoris), we further coloured them according to the limb they play a part in moving. On the left, the four colours are described with a legend: left upper limb (blue), right upper limb (green), left lower limb (yellow), and right lower limb (red).}
    \label{fig:kinesismusclemodelandMuscleGroups}
\end{figure}

The different coloured areas on the avatar are representing the following muscle groups: M. Deltoideus, M. Pectoralis Major, M. Triceps Brachii, M. Biceps Brachii and M. Brachialis, M. Latissimus Dorsi, M. Gluteus Maximus, Mm. ischiocrurales, M. Quadriceps Femoris \autoref{fig:kinesismusclemodelandMuscleGroups}. 
The higher the usage of the muscle group, the higher the colour on the gradient scale from turquoise to red.  
For a less accurate but more comparable representation, the box plots only show even higher-level muscle groups, the left upper limb (blue), left lower limb (green), left lower limb (yellow), right lower limb (red) and overall value (white).
% wie funktioniert / steigt die scala : like (i.e., more deviation from the baseline ranging from -1.0 (maximum contracted) to 1.0 (maximum stretched)).

\subsection{Implementation}

\begin{figure}[!t]
    \centering
    \includegraphics[width=0.5\textwidth]{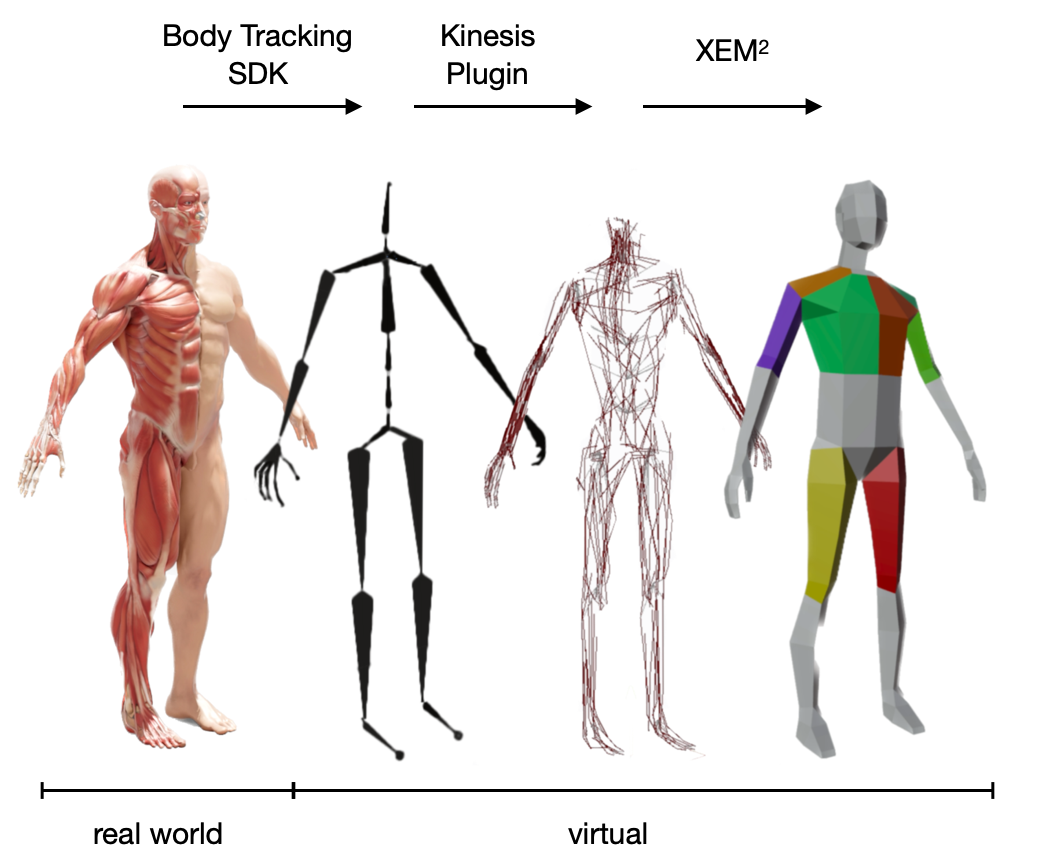} 
    \caption{This Figure represents the implementation stack of the process from the movement to the \EXMA visualization. On the left, there is a real-world human executing the movement. With the help of the Body Tracking SDK, the skeleton of the movement is transferred into a virtual model in Unity. In Unity, the Kinesis plugin transfers the movement into a muscle model where the forces of muscles are calculated and visualized as groups in the last highlighted 3D human model through \EXMA.}
    \Description{The figure shows 4 avatars in a row. The first avatar is a 3D human muscle model labelled underneath with the real world. Underneath the other three avatars, it says virtual. From the first avatar, there is an arrow to the second avatar, saying Body Tracking SDK. The second avatar is a 3D model skeleton with an arrow to the next avatar saying Kinesis Plugin. The third avatar shows the muscles as lines from over 200 muscles of the Kinesis muscle model. From the third avatar, there is an arrow to the fourth avatar saying XEM$^2$. The third avatar shows the human model we used to visualize the muscle work of different muscle groups.}
    \label{fig:implstack}
\end{figure}

This section describes the details of the \EXMA~ system implementation, starting from the motion tracking SDK, the Kinesis muscle model asset for calculating algorithms of the muscle work using the Hill-type muscle model to the user visualization of the muscle work. The implementation and model stack \EXMA~ is based on can be seen in \autoref{fig:implstack}.

\subsubsection{Motion Tracking \& Unity}
We use the Motion Tracking SDK (V. 1.1.2) from the Azure Kinect to get the skeleton movement data into Unity\footnote{\url{https://unity.com/}} (V. 2021.3.2f1) for visualization and in-game adaption. 
Using the Azure Kinect examples for Unity (V. 1.17.3) as a foundation, we import motion data into Unity. After installing and setting up all the requirements for the Body Tracking SDK and the Unity plug-in we can see only the skeleton of a virtual avatar moving. 

\subsubsection{Hill-type Muscle Model}
The Hill-type muscle model is a biomechanical model that describes the behaviour of skeletal muscles~\cite{zhouRoughPreciseHumanInspired2019}. 
It ignores internal processes in the muscles and instead sees the muscle as a contractile that needs work effort to contract. 

The Hill-type muscle model assumes that a muscle can be represented as a contractile element (CE) in parallel with a passive elastic element (PE). 
The CE generates force by shortening or lengthening, while the PE resists changes in length. 
In addition, the model includes a series of elastic elements (SE) that are attached to the tendon of the muscle.

The force generated by the CE is modelled using a force-velocity relationship that takes into account the velocity at which the muscle is contracting or lengthening. 
The maximum force that the CE can generate is proportional to its physiological cross-sectional area, which is a measure of the muscle's force-generating capacity. 
The force generated by the PE is modelled using a passive force-length relationship that describes the stiffness of the muscle at different lengths.

The force generated by the muscle as a whole is the sum of the forces generated by the CE and the PE. 
The SE acts as a spring, storing and releasing energy as the muscle contracts and lengthens.

%input: muscle load, muscle length, and muscle stimulation
The underlying equation for the CE force-velocity looks as follows for a shortening muscle \cite{wintersHillbasedMuscleModels1990}:

\begin{equation}
  (F + a)*(v + b) = (F_{max} + a)*b,
\end{equation}

where $F$ is the tension of the muscle, $v$ the velocity of the contraction, $F_{max}$ the maximum isometric tension, and $a$ and $b$ are constants. 

The Hill-type muscle model is often used to simulate muscle behaviour in different conditions, such as during different types of muscle contractions or during fatigue. It has been widely used in biomechanics research~\cite{cailletHilltypeComputationalModels2022, schmittDynamicsSkeletalMuscle2019} and in the design of prosthetic limbs~\cite{eilenbergControlPoweredAnkle2010} and other assistive devices~\cite{zhouDesignResearchLower2020}.
We chose this representation for the muscles because it is most practical to work with in a digital simulation. The forces of each individual muscle can be extracted and worked with without any conversion, which would be prone to reinforcing inaccuracies. 

\subsubsection{Kinesis -- Physical Muscle Model}
We used the \emph{Kinesis - Physical Muscle Model Based Movement (V. 1.0.0)}\footnote{https://assetstore.unity.com/packages/tools/physics/kinesis-physical-muscle-model-based-movement-206089} Unity plugin as muscle model for our calculations. 
The Kinesis Muscle Model has over 200 individual muscles arranged in a human skeleton model that quite closely imitates the real human body. 
Nevertheless, in human anatomy, there are individual elevations and structures, various ligaments or individual muscle lengths which ``redirect'' the muscles and their power and affect muscle work. 
This is missing in Kinesis or is rather primitively reproduced so that the movement and calculation themselves work but on a generic body model without including any individual refinements. 

Kinesis is built to manipulate different muscles and then calculate the resulting force with the Hill-Type Model and animate the avatar.
Therefore, we inverted the processing pipeline for our purpose. 
We used the camera-tracked skeleton motion, to move and animate the Kinesis Model. 
Through the animation, the simulated muscles were moved as contraction or stretching and we reversed the calculation to match the force needed to move the skeleton to the target position and hold the position accordingly.  

%Through the movement one could then read out as a kind of reversal of actio/reactio in the simulated muscles, which force these would have to counteract in order to hold the current position of the skeleton. Thus, as said above, also the aspect of concentric work could not be calculated, but only the current force that the muscle must apply, quasi for a kind of snapshot.

While the Kinesis is based on the work of \citet{geijtenbeekFlexibleMusclebasedLocomotion2013}, we further expanded and adjusted it to operate smoothly in Unity. 
The core structure consists of three parts, the muscle node (bones to which muscles are attached), the muscle segment (rigid body, parentage, head, tail, and joints) and the muscle tension unit (muscle as a whole in a more technical way).  
The Kinesis model only includes muscle strings with default length. Details like height, BMI (body mass index), etc., are not included in the measurement (see Hill-Type Model). 

In our case, we extract the force F that Unity applies to individual muscles at any given time.
This value is used to compute the relation of the movement of individual body parts to the total movement of all muscles.
Based on the division into functional muscle groups, we can compute a relative value of forces, which can then be translated into movement or even exertion given a reference value. 
Since visualizing all muscles would clutter the scene and, therefore, would not be very helpful for non-expert users, we grouped individual muscles into muscle groups (summing them up, no interpolation).
To define the muscle groups used in \EXMA, we used the Anatomy Atlas~\cite{rossThiemeAtlasAnatomy2006}.
Specifically, we based our grouping on prime movers for the upper and lower limbs defined in the Anatomy Atlas~\cite{rossThiemeAtlasAnatomy2006} to simplify the visualization and computation efforts.
The results of the muscle forces were either visualized in bar charts or on an abstract human model.

\subsubsection{User Visualization Screen}

For the 3D highlighted human model, we used the median of accumulated values over the last second (60fps) and coloured the associated muscle group. The colour is chosen between a value from zero (least movement) to the maximum highest accumulated total value and use it to define the 100\% (most movement).
For the box plots, we grouped the muscle group values again into the four limbs \autoref{fig:kinesismusclemodelandMuscleGroups} and visualized the value directly as the height of the chart.

\begin{figure}[h]
    \centering
    \includegraphics[width=0.5\textwidth]{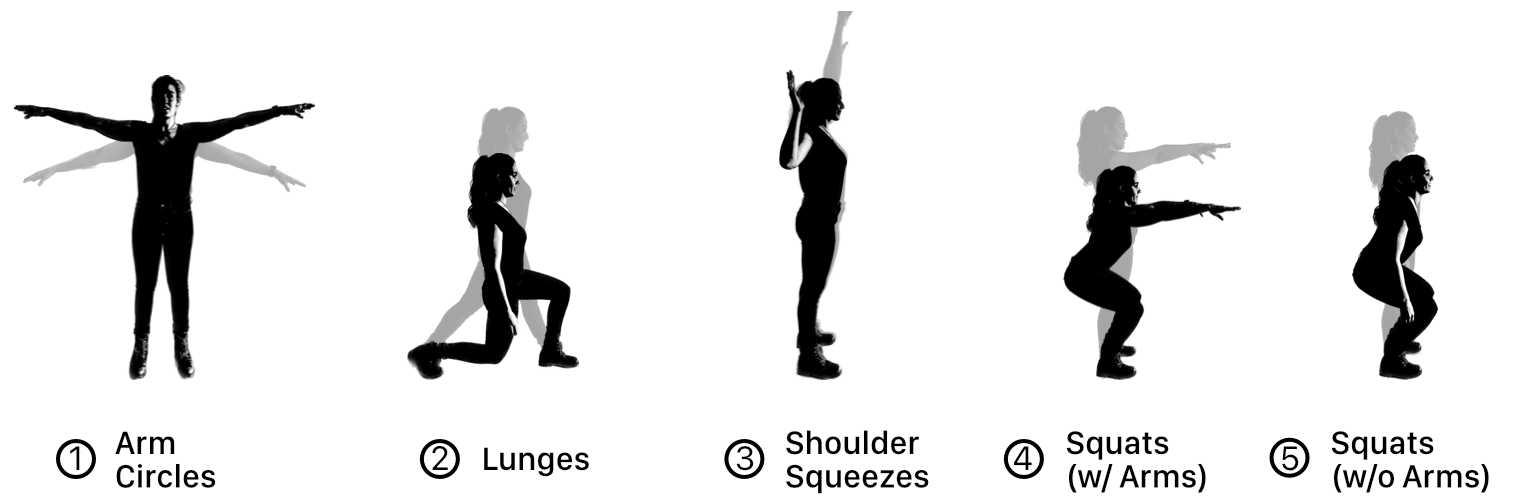}
    \caption{The five exercises that were used to measure muscle work with \EXMA for technical validity. Each exercise was executed 5 times in one measurement and five measurements were taken in total. Each measurement consisted of 5 executions of a movement, yielding a total of $5x5x5=125$ sampled motions.}
    \Description{The figures shows five exercises from left to right: (1) arms are spread to the side and make circles. (2) lunges, one foot is front, one is back and the body moves up and down between long legs and legs in 90 degree. (3) shoulder pushes, the arms are stretched upwards and pulled down at the elbow while the shoulder plates squeeze together. (4) squats with arm stretched towards the front and hold there the whole time. (5) squats with arms hanging sideways down.}
    \label{fig:plottasks}
\end{figure}

\section{Technical Validity}
To assess the reliability of \EXMA's measurements, we conducted a technical validation study with N=10 people and five exercises to show how the values change for each muscle group and person when doing an exercise repeatedly.
A valid measurement would show only slightly different results for each participant since each person has slightly different muscle lengths and muscle body compositions. 
The value changes for each muscle group should differ according to the muscles that are used for this specific exercise.

\subsection{Method}
We decided on five exercises, shown in \autoref{fig:plottasks}.
The set consists of two exercises for the arms (\textcircled{1} arm circles and \textcircled{3}
 shoulder squeezes), two exercises for the legs (\textcircled{4} squat with arms \textcircled{5} squats without arms), and one exercise with an asymmetric load on the legs (\textcircled{2} lunges).
All 10 participants were doing five measurements with each a set of five for each exercise: 10 users x 5 exercises x 5 measurements x 5 repetitions. One measurement had a set of 5 exercises to smooth out natural differences in movement and avoid noise.
%This approach is particularly relevant for shorter measurements.
The code for the technical validation is slightly different than from the usability study, since we found a small error in the implementation.
More details are in the limitations section. 

\subsection{Results}

\begin{figure*}[h]
    \centering
    \includegraphics[width=1.00\textwidth, trim=4em 6em 4em 4em,clip]{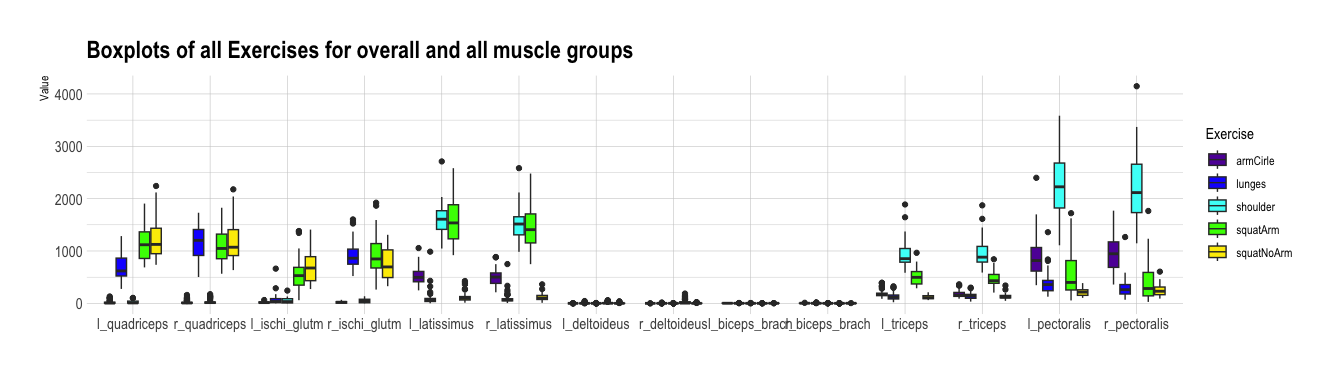}
    \caption{Plotted visualization of five exercises (arm circles, lunges, shoulder squeeze, squats with arms, squats without arms) measurements of \EXMA, showing values separately for each muscle group.}
    \Description{Plotted visualization of five exercises (arm circles, lunges, shoulder squeeze, squats with arms, squats without arms) measurements of \EXMA, showing separately for each muscle group. Obvious are the strong differences between lunges l./r. ischiocrurales.}
    \label{fig:plottechvalid}
\end{figure*}

% NEW TABLE

\begin{table*}[h]
\caption{The table shows for each muscle group the significances of the repeated measures ANOVA for exercises, user IDs, and the interaction between both. For clarity, cells with no significant results are left blank.}
\label{tab:festability}
\resizebox{\textwidth}{!}{%
\renewcommand{\arraystretch}{1.2}
\begin{tabular}{clllllllllllllll}
~ &
   &
  \rotatebox{0}{\textbf{l. quad}} &
  \rotatebox{0}{\textbf{r. quad}} &
  \rotatebox{0}{\textbf{l. ischi}} &
  \rotatebox{0}{\textbf{r. ischi}} &
  \rotatebox{0}{\textbf{l. lat}} &
  \rotatebox{0}{\textbf{r. lat}} &
  \rotatebox{0}{\textbf{l. del}} &
  \rotatebox{0}{\textbf{r. del}} &
  \rotatebox{0}{\textbf{l. bic}} &
  \rotatebox{0}{\textbf{r. bic}} &
  \rotatebox{0}{\textbf{l. tric}} &
  \rotatebox{0}{\textbf{r. tric}} &
  \rotatebox{0}{\textbf{l. pec}} &
  \rotatebox{0}{\textbf{r. pec}} \\ \hline
\multicolumn{1}{c|}{} &
  \multicolumn{1}{l|}{\cellcolor[HTML]{EFEFEF}\textbf{F(240)}} &
  \multicolumn{1}{l|}{\cellcolor[HTML]{EFEFEF}325.876} &
  \multicolumn{1}{l|}{\cellcolor[HTML]{EFEFEF}296.235} &
  \multicolumn{1}{l|}{\cellcolor[HTML]{EFEFEF}135.825} &
  \multicolumn{1}{l|}{\cellcolor[HTML]{EFEFEF}193.033} &
  \multicolumn{1}{l|}{\cellcolor[HTML]{EFEFEF}472.093} &
  \multicolumn{1}{l|}{\cellcolor[HTML]{EFEFEF}421.257} &
  \multicolumn{1}{l|}{\cellcolor[HTML]{EFEFEF}12.204} &
  \multicolumn{1}{l|}{\cellcolor[HTML]{EFEFEF}15.893} &
  \multicolumn{1}{l|}{\cellcolor[HTML]{EFEFEF}18.732} &
  \multicolumn{1}{l|}{\cellcolor[HTML]{EFEFEF}12.060} &
  \multicolumn{1}{l|}{\cellcolor[HTML]{EFEFEF}338.795} &
  \multicolumn{1}{l|}{\cellcolor[HTML]{EFEFEF}350.445} &
  \multicolumn{1}{l|}{\cellcolor[HTML]{EFEFEF}250.101} &
  \multicolumn{1}{l|}{\cellcolor[HTML]{EFEFEF}254.314} \\
\multicolumn{1}{c|}{} &
  \multicolumn{1}{l|}{\textbf{p <}} &
  \multicolumn{1}{l|}{.0001} &
  \multicolumn{1}{l|}{.0001} &
  \multicolumn{1}{l|}{.0001} &
  \multicolumn{1}{l|}{.0001} &
  \multicolumn{1}{l|}{.0001} &
  \multicolumn{1}{l|}{.0001} &
  \multicolumn{1}{l|}{.0001} &
  \multicolumn{1}{l|}{.0001} &
  \multicolumn{1}{l|}{.0001} &
  \multicolumn{1}{l|}{.0001} &
  \multicolumn{1}{l|}{.0001} &
  \multicolumn{1}{l|}{.0001} &
  \multicolumn{1}{l|}{.0001} &
  \multicolumn{1}{l|}{.0001} \\
\multicolumn{1}{c|}{\multirow{-3}{*}{\rotatebox{90}{\textbf{Exercise}\hspace{0.2em}}}} &
  \multicolumn{1}{l|}{\cellcolor[HTML]{EFEFEF}\textbf{$\eta^{2}$}} &
  \multicolumn{1}{l|}{\cellcolor[HTML]{EFEFEF}0.845 (l)} &
  \multicolumn{1}{l|}{\cellcolor[HTML]{EFEFEF}0.832 (l)} &
  \multicolumn{1}{l|}{\cellcolor[HTML]{EFEFEF}0.694 (m)} &
  \multicolumn{1}{l|}{\cellcolor[HTML]{EFEFEF}0.763 (m)} &
  \multicolumn{1}{l|}{\cellcolor[HTML]{EFEFEF}0.887 (l)} &
  \multicolumn{1}{l|}{\cellcolor[HTML]{EFEFEF}0.875 (l)} &
  \multicolumn{1}{l|}{\cellcolor[HTML]{EFEFEF}0.169} &
  \multicolumn{1}{l|}{\cellcolor[HTML]{EFEFEF}0.209 (s)} &
  \multicolumn{1}{l|}{\cellcolor[HTML]{EFEFEF}0.238 (s)} &
  \multicolumn{1}{l|}{\cellcolor[HTML]{EFEFEF}0.167} &
  \multicolumn{1}{l|}{\cellcolor[HTML]{EFEFEF}0.850 (l)} &
  \multicolumn{1}{l|}{\cellcolor[HTML]{EFEFEF}0.854 (l)} &
  \multicolumn{1}{l|}{\cellcolor[HTML]{EFEFEF}0.807 (l)} &
  \multicolumn{1}{l|}{\cellcolor[HTML]{EFEFEF}0.809 (l)} \\ \hline
\multicolumn{1}{c|}{} &
  \multicolumn{1}{l|}{\textbf{F(240)}} &
  \multicolumn{1}{l|}{16.152} &
  \multicolumn{1}{l|}{25.353} &
  \multicolumn{1}{l|}{4.986} &
  \multicolumn{1}{l|}{~} &
  \multicolumn{1}{l|}{~} &
  \multicolumn{1}{l|}{5.518} &
  \multicolumn{1}{l|}{4.103} &
  \multicolumn{1}{l|}{~} &
  \multicolumn{1}{l|}{9.862} &
  \multicolumn{1}{l|}{~} &
  \multicolumn{1}{l|}{15.525} &
  \multicolumn{1}{l|}{17.583} &
  \multicolumn{1}{l|}{12.657} &
  \multicolumn{1}{l|}{8.820} \\
\multicolumn{1}{c|}{} &
  \multicolumn{1}{l|}{\cellcolor[HTML]{EFEFEF}\textbf{p <}} &
  \multicolumn{1}{l|}{\cellcolor[HTML]{EFEFEF}.0001} &
  \multicolumn{1}{l|}{\cellcolor[HTML]{EFEFEF}.002} &
  \multicolumn{1}{l|}{\cellcolor[HTML]{EFEFEF}.05} &
  \multicolumn{1}{l|}{\cellcolor[HTML]{EFEFEF}~} &
  \multicolumn{1}{l|}{\cellcolor[HTML]{EFEFEF}~} &
  \multicolumn{1}{l|}{\cellcolor[HTML]{EFEFEF}.05} &
  \multicolumn{1}{l|}{\cellcolor[HTML]{EFEFEF}0.05} &
  \multicolumn{1}{l|}{\cellcolor[HTML]{EFEFEF}~} &
  \multicolumn{1}{l|}{\cellcolor[HTML]{EFEFEF}.005} &
  \multicolumn{1}{l|}{\cellcolor[HTML]{EFEFEF}~} &
  \multicolumn{1}{l|}{\cellcolor[HTML]{EFEFEF}.0002} &
  \multicolumn{1}{l|}{\cellcolor[HTML]{EFEFEF}.0001} &
  \multicolumn{1}{l|}{\cellcolor[HTML]{EFEFEF}.0001} &
  \multicolumn{1}{l|}{\cellcolor[HTML]{EFEFEF}.005} \\
\multicolumn{1}{c|}{\multirow{-3}{*}{\rotatebox{90}{\textbf{ID}}}} &
  \multicolumn{1}{l|}{\textbf{$\eta^{2}$}} &
  \multicolumn{1}{l|}{0.063} &
  \multicolumn{1}{l|}{0.096} &
  \multicolumn{1}{l|}{0.02} &
  \multicolumn{1}{l|}{~} &
  \multicolumn{1}{l|}{~} &
  \multicolumn{1}{l|}{0.022} &
  \multicolumn{1}{l|}{0.017} &
  \multicolumn{1}{l|}{~} &
  \multicolumn{1}{l|}{0.039} &
  \multicolumn{1}{l|}{~} &
  \multicolumn{1}{l|}{0.061} &
  \multicolumn{1}{l|}{0.068} &
  \multicolumn{1}{l|}{0.050} &
  \multicolumn{1}{l|}{0.035} \\ \hline
\multicolumn{1}{c|}{} &
  \multicolumn{1}{l|}{\cellcolor[HTML]{EFEFEF}\textbf{F(240)}} &
  \multicolumn{1}{l|}{\cellcolor[HTML]{EFEFEF}3.321} &
  \multicolumn{1}{l|}{\cellcolor[HTML]{EFEFEF}2.609} &
  \multicolumn{1}{l|}{\cellcolor[HTML]{EFEFEF}~} &
  \multicolumn{1}{l|}{\cellcolor[HTML]{EFEFEF}2.805} &
  \multicolumn{1}{l|}{\cellcolor[HTML]{EFEFEF}~} &
  \multicolumn{1}{l|}{\cellcolor[HTML]{EFEFEF}2.483} &
  \multicolumn{1}{l|}{\cellcolor[HTML]{EFEFEF}3.993} &
  \multicolumn{1}{l|}{\cellcolor[HTML]{EFEFEF}~} &
  \multicolumn{1}{l|}{\cellcolor[HTML]{EFEFEF}~} &
  \multicolumn{1}{l|}{\cellcolor[HTML]{EFEFEF}~} &
  \multicolumn{1}{l|}{\cellcolor[HTML]{EFEFEF}20.45} &
  \multicolumn{1}{l|}{\cellcolor[HTML]{EFEFEF}4.638} &
  \multicolumn{1}{l|}{\cellcolor[HTML]{EFEFEF}7.248} &
  \multicolumn{1}{l|}{\cellcolor[HTML]{EFEFEF}~} \\
\multicolumn{1}{c|}{} &
  \multicolumn{1}{l|}{\textbf{p <}} &
  \multicolumn{1}{l|}{.05} &
  \multicolumn{1}{l|}{.05} &
  \multicolumn{1}{l|}{~} &
  \multicolumn{1}{l|}{.05} &
  \multicolumn{1}{l|}{~} &
  \multicolumn{1}{l|}{.05} &
  \multicolumn{1}{l|}{0.005} &
  \multicolumn{1}{l|}{~} &
  \multicolumn{1}{l|}{~} &
  \multicolumn{1}{l|}{~} &
  \multicolumn{1}{l|}{.05} &
  \multicolumn{1}{l|}{.002} &
  \multicolumn{1}{l|}{.0001} &
  \multicolumn{1}{l|}{~} \\
\multicolumn{1}{c|}{\multirow{-3}{*}{\rotatebox{90}{\textbf{Exer:ID }}}} &
  \multicolumn{1}{l|}{\cellcolor[HTML]{EFEFEF}\textbf{$\eta^{2}$}} &
  \multicolumn{1}{l|}{\cellcolor[HTML]{EFEFEF}0.052} &
  \multicolumn{1}{l|}{\cellcolor[HTML]{EFEFEF}0.042} &
  \multicolumn{1}{l|}{\cellcolor[HTML]{EFEFEF}~} &
  \multicolumn{1}{l|}{\cellcolor[HTML]{EFEFEF}0.045} &
  \multicolumn{1}{l|}{\cellcolor[HTML]{EFEFEF}~} &
  \multicolumn{1}{l|}{\cellcolor[HTML]{EFEFEF}0.040} &
  \multicolumn{1}{l|}{\cellcolor[HTML]{EFEFEF}0.062} &
  \multicolumn{1}{l|}{\cellcolor[HTML]{EFEFEF}~} &
  \multicolumn{1}{l|}{\cellcolor[HTML]{EFEFEF}~} &
  \multicolumn{1}{l|}{\cellcolor[HTML]{EFEFEF}~} &
  \multicolumn{1}{l|}{\cellcolor[HTML]{EFEFEF}0.052} &
  \multicolumn{1}{l|}{\cellcolor[HTML]{EFEFEF}0.072} &
  \multicolumn{1}{l|}{\cellcolor[HTML]{EFEFEF}0.108} &
  \multicolumn{1}{l|}{\cellcolor[HTML]{EFEFEF}~} \\ \hline
\end{tabular}}
\end{table*}

 We statistically evaluated the measurements using repeated measures ANOVA.

%\paragraph{Accuracy}
%To validate if the measurement for each person and each exercise is comparable, we compared measurements for each participant separately. 
%\textbf{\Huge\color{red}TODO}

%\paragraph{Precision}
To investigate precision, we compared the means of the exercises and users (\autoref{tab:festability}). We found significant differences between every muscle group and exercise with medium to high effect size. This result indicates that the measurement is precise enough to differentiate between the exercises for the muscle groups. 
However, post-hoc test show that some of the muscle groups are not significantly different from each other, which makes sense for movements that use the same muscle group.

When looking into post-hoc tests (pairwise comparison with Tukey HSD, we find many significant results between exercises for l. quadriceps, r. quadriceps, l. ischi. glutm, r. ischi. glutm, l. latissimus,r. latissimus l. deltoideus, r. deltoideus, l. biceps brach., r. biceps brach., l. triceps, r. triceps, l. pectoralis, r. pectoralis (details can be found in the supplementary materials).
We find arm circles and shoulder squeezes as not significantly different, often in the leg muscle groups, while non-significances between lunges and squats are often found for arm and shoulder muscle groups.

Regarding the users, the significances when comparing between muscle groups are rather low (effect size is negligible), which means that values can be quite similar between users. This hints at the notion that different body sizes and forms do not make too much of a practical difference in measuring muscle work. 
Measurement results are shown as boxplots in \autoref{fig:plottechvalid}.

Observing the boxplots and pairwise comparisons, we can see that for squats with and without arms the l. and r. pectoralis and l. and r. triceps muscles means significantly differ, meaning that the system detected correctly when the arms where used and when not.
The same validation can be seen for lunges and squads. The means are almost similar for l. and r. quadriceps and l. and r. ischi glutm. when doing squats, but for lunges ,there is an obvious difference between l. and r. quadriceps and l. and r. ischi glutm. . This is because the leg muscle groups are used differently in the asymmetric lunge exercise.

%While the measures are not perfectly the same due to a certain degree of variance of movement or camera tracking angle, we see that the values of active muscles are within the same range. More precisely, we see that when arm muscles were inactive, the values our system calculated for arm muscles (l./r. latissimus, l./r. triceps, l./r. pectoralis) are low compared to the instances of active arm muscles.
%That the l./r. latissimus dorci has such a huge impact when holding up the arms against leaving them next to the body,
Fluctuations can occur due to tracking inaccuracies or when the movement is not executed very precisely. Specifically, we observed glitches with the arm tracking while doing this technical validation. 
This is very likely causing the higher values for arm muscle groups compared to the leg muscle groups. However, the ratios between muscles and between the leg muscle groups and the arm muscle groups are still fitting.

\section{User Study}\label{lab:userstudy}
We evaluated the acceptance, usefulness, benefits, and opportunities for improvement of \EXMA~with a within-participant study. %consisting of two parts.

The goal of the study was to understand the different measurement results of \EXMA, BC and RPE and how users evaluated our approach compared to existing measures (BC and RPE) that users were likely more familiar with.
%The evaluation goal of the second part was only to focus on the visualization of the tool itself. The idea here was to investigate how the visualization during the workout would influence users' perceived competence, usefulness, and awareness.

%\begin{figure*}[h]
%    \centering
%    \includegraphics[width=1.00\textwidth]{images/ablauf.png}
%    \caption{Procedure of the User Study including both parts, with approximate durations shown above each step.}
%    \Description{The process of the user study starts with 10 minutes of Consent, Introduction and Preparation of the hardware. Then there are 20 minutes of study part one including FitXR Tutorial and one Level. Afterwards three times they were presented with an exertion measurement result and had to fill out a survey. Next was study part two with 20 minutes, where they played and large screen exergame two times and filled out a survey afterwards. Finally, in the last 10 minutes, a demographics questionnaire was filled out and participants got remuneration.}
%    \label{fig:studyprocess}
%\end{figure*}

%We focused on measurement differences and participants' perception of \EXMA~ compared compared to existing measures (BC and RPE) that users were likely more familiar with. %to other exhaustion measurement techniques.
%The second part focused on the visualization of muscle work itself using \EXMA. 
%In particular, we evaluated the difference between in-situ visualization to no visualization. %in a small custom-designed exergame. 
The study used an exergame, FitXR\footnote{\url{https://fitxr.com/}}, as a use-case, as there are no generic precise measurement methods for exergames and there may be an imbalance between different muscle groups. 
%The first part was conducted with a virtual reality (VR) Headset and the second part was conducted using a large screen, to show the variability and independence of the presentation tool.    

\subsection{Participants}

We recruited 36 participants (16 female, 21 male, 0 non-binary,  M$_{age}$ = 24.95, SD = 4.98, Min-Max= 18--42). 
78.38\% of the participants were university students and 21.62\% were employed. 
Over 50\% stated that they exercise frequently or almost daily and only one person stated to never do physical exercises.
89.19\% stated that they had never used the commercial FitXR exergame and 70.27\% had never used a VR headset before. 

\begin{figure}[h]
    \centering
    \includegraphics[width=0.5\textwidth]{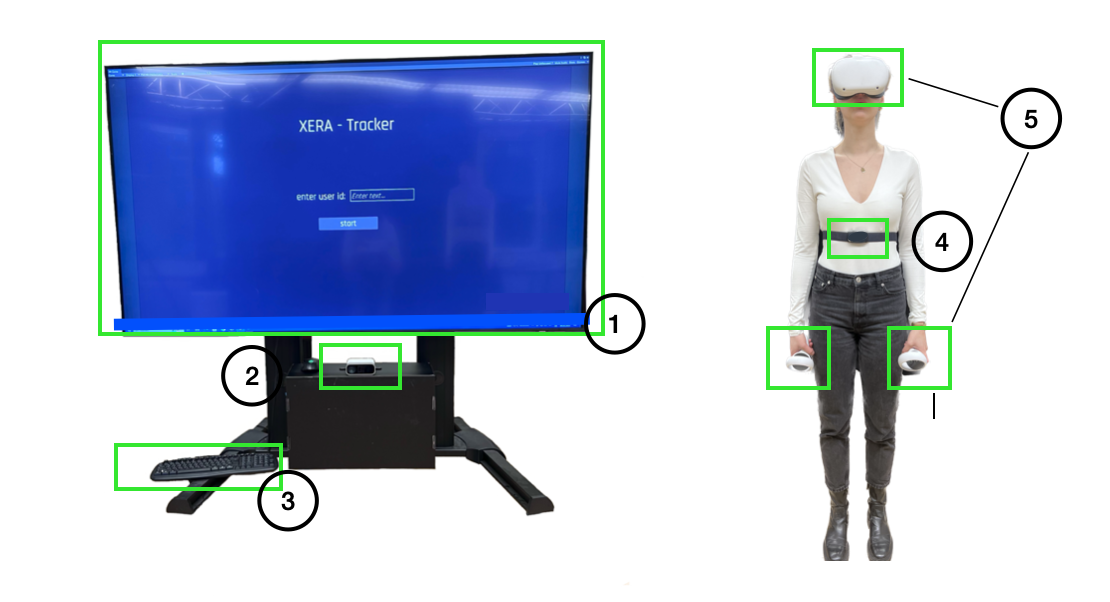}
    \caption{The hardware setup for the study. On the left side is the 75 inches visualization screen \textcircled{1}, the Azure Kinect Camera \textcircled{2}, and a keyboard for text input \textcircled{3}. The right side depicts the participants' instrumentation: a participant wearing a Polar H10 heart rate sensor \textcircled{4} and the Oculus Quest 2 headset, both for study part one \textcircled{5}.}
    \Description{On the left side there is a large 75-inch screen labelled with a circled one. Below the screen, there is an Azure Kinect with a circled two and a keyboard on the side with a circled three. On the right side, there is a participant wearing the polar heart rate sensor, labelled with number four and an oculus quest 2 headset labelled with number five.}
    \label{fig:setup}
\end{figure}

\subsection[Comparing Exertion Feedback]{Comparing Exertion Feedback Using RPE-Scale, Burned Calories and \EXMA} 

Since burned calories \cite{ainsworthCompendiumPhysicalActivities2000} are a common technique for measuring exertion \cite{mopasTrainingFeelWearable2020} and the RPE scale \cite{williamsBorgRatingPerceived2017} an established self-perception measurement for athletes, we decided to use both for the perception comparison study part. 
The first study part was guided by the following research question: 

%For the evaluation of muscle work with \EXMA compared to burned calories and RPE Scale\cite{} in the first within-participant study part 

\emph{RQ: How does usefulness, motivation, trust, familiarity, body understanding, and perceived precision differ between physical exercise measures of the RPE scale, burned calories and muscle work measured with the \EXMA~system?}

\subsubsection{Setup}
The participants wore a Meta Quest 2 headset (\autoref{fig:setup} \textcircled{5}) and a heart rate sensor (\autoref{fig:setup} \textcircled{4}). 
We used the tutorial and first boxing level from FitXR\footnote{\url{https://www.oculus.com/experiences/quest/2327205800645550/}} as a workout. 
In this level, participants had to hit coloured spheres with their hands and dodge objects by squatting or leaning to the left or right. 
%It took about X minutes to finish the tutorial and X to finish the first level.

\subsubsection{Conditions}

This first study part consisted of three conditions that were counterbalanced between the participants. 
After a FitXR exercise session, participants were confronted with three different measurement methods: 
%In each of the three conditions, the participants first played one level of the FitXR exergame in VR and then assessed their level of exertion with one of three measurement methods: (1) the RPE scale (2), burned calories, (3) \EXMA.
%We evaluated three measurement techniques for physical exercise in the first study (two known ones and one new one, our \EXMA system), after they performed the FitXR exergame in virtual reality. 

\begin{figure}[h]
    \centering
    \includegraphics[width=0.5\textwidth]{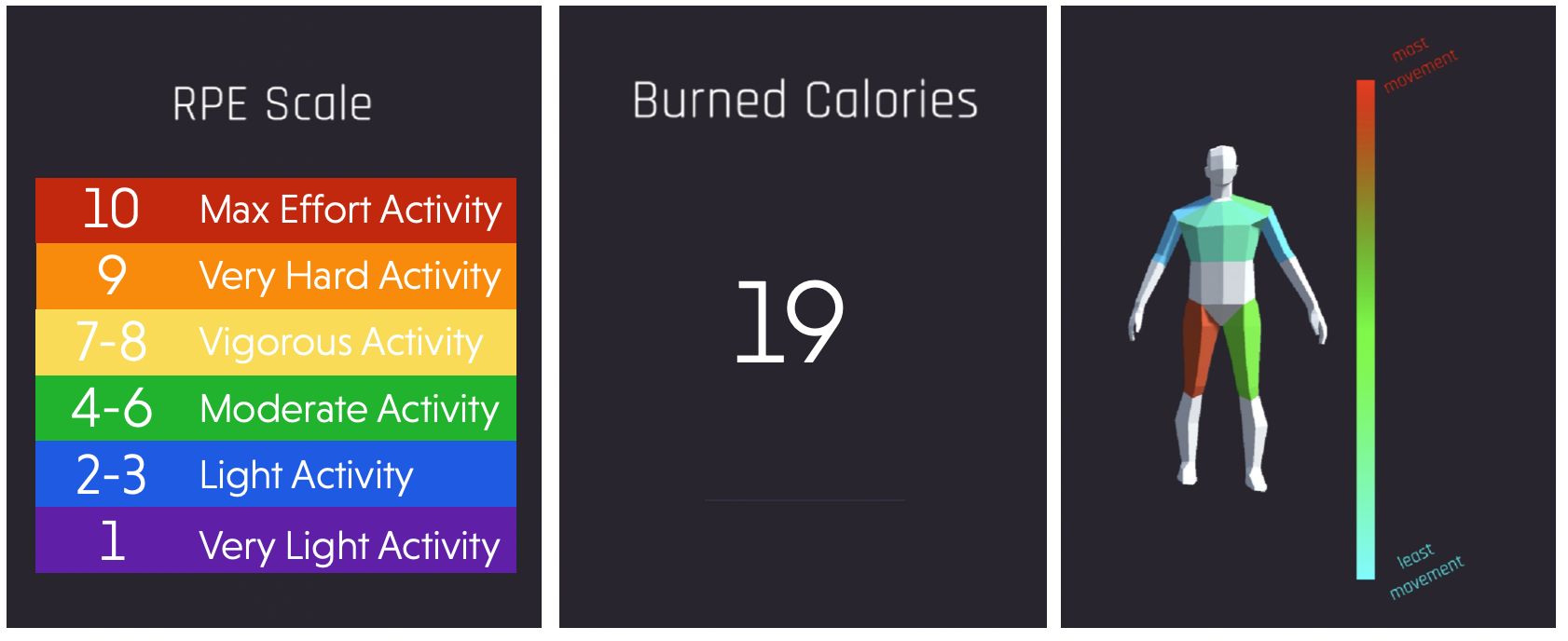}
    \caption{On the left, we show the self-perception RPE Scale using 10 Levels. In the middle, we show the calculated burned calories value, and on the right, there is the relative muscle work by each muscle group visualized on a 3D Avatar.}
    \Description{The figure shows three different exertion techniques. First the 10-level RPE scale: 1 = very light activity, 2-3 = light activity, 4-6 = moderate activity, 7-8 = vigorous activity, 9 = very hard activity, and 10 max effort activity. In the middle there is the value for burned calories and on the right, there is the highlighted avatar, with colour gradient, depending on the level of exertion for the muscle group.}
    \label{fig:conditions1}
\end{figure}

\begin{itemize}

\item[(1)] \textit{Perceived Exertion-Scale (RPE):} Participants had to evaluate their physical exertion level using the RPE scale. 
They were shown the RPE scale with Levels from 1 to 10 and short explanations about the exertion of this level. Then they had to pick their own level in the scale and enter it using the keyboard. (\autoref{fig:conditions1}, left). Then, they were given information on whether the measured heart rate was equal, more, or less than the RPE value converted to heart rate.
For analysis we used the self-reported RPE value. 

\item[(2)]\textit{Burned Calories (BC):} Participants were shown their level of burned calories (in kcal) measured and calculated using a heart rate sensor they were wearing on their upper body (\autoref{fig:conditions1}, middle). Additionally, they were given information on the equivalent of food calories they burned in this session. 

\item[(3)]\textit{Muscle-work measured with \EXMA (MW):} We showed the participants our visualisation of muscle work. 
They could see a bar chart with overall and prime mover (left/right upper/lower limb) and an avatar showing which muscle group was used and how much indicated by a colour gradient (\autoref{fig:conditions1}, right).   
\end{itemize}

\subsubsection{Measures}

\paragraph{Heart rate}
Heart rate (average and peak) was measured with the Polar Heart Rate Sensor H10\footnote{https://www.polar.com/en/sensors/h10-heart-rate-sensor}. 
Each participant wore it around their lower chest while playing. 
As an exercise type, we chose ``other indoor'' as exercise on the heart rate sensor since there was no other suitable value.

\paragraph{Physical Activity Measures}
Besides the accumulated values of \EXMA~ overall and for all muscle groups separately, we logged anonymized user input: age, weight, and heart rate from the sensor to calculate and log burned calories. 
Lastly, the user input of the chosen value from the RPE Scale was saved.
Based on a conversion of the RPE scale to a heart rate value, we stored and displayed whether the converted heart rate differs from the measured heart rate (i.e., lower, equal, or higher).

\paragraph{Post-Condition Questions}
After each condition, we asked participants 7-point-Likert questions (1= strongly disagree to 7= strongly agree) about usefulness, trust, expected value, familiarity with the method, current use and future use\footnote{Asking participants if they currently use similar measurement techniques like PRE, BC, MW (current use) and if they could imagine using one of them in the future (future use).}, body understanding, and precision. 
Depending on the conditions we asked, if it was hard to rate using the RPE Scale, if they know how burned calories are calculated, and if they preferred a different representation of muscle work (textual or as a single value). 
Lastly, participants could write some advantages and disadvantages they encountered.

Post all three conditions we further asked participants to rank the three conditions according to usefulness, trust, future use, most information, motivation for future workouts, and whether two of the measurements should be combined.  

\begin{figure*}[!t]
    \centering
    \includegraphics[width=1.00\textwidth]{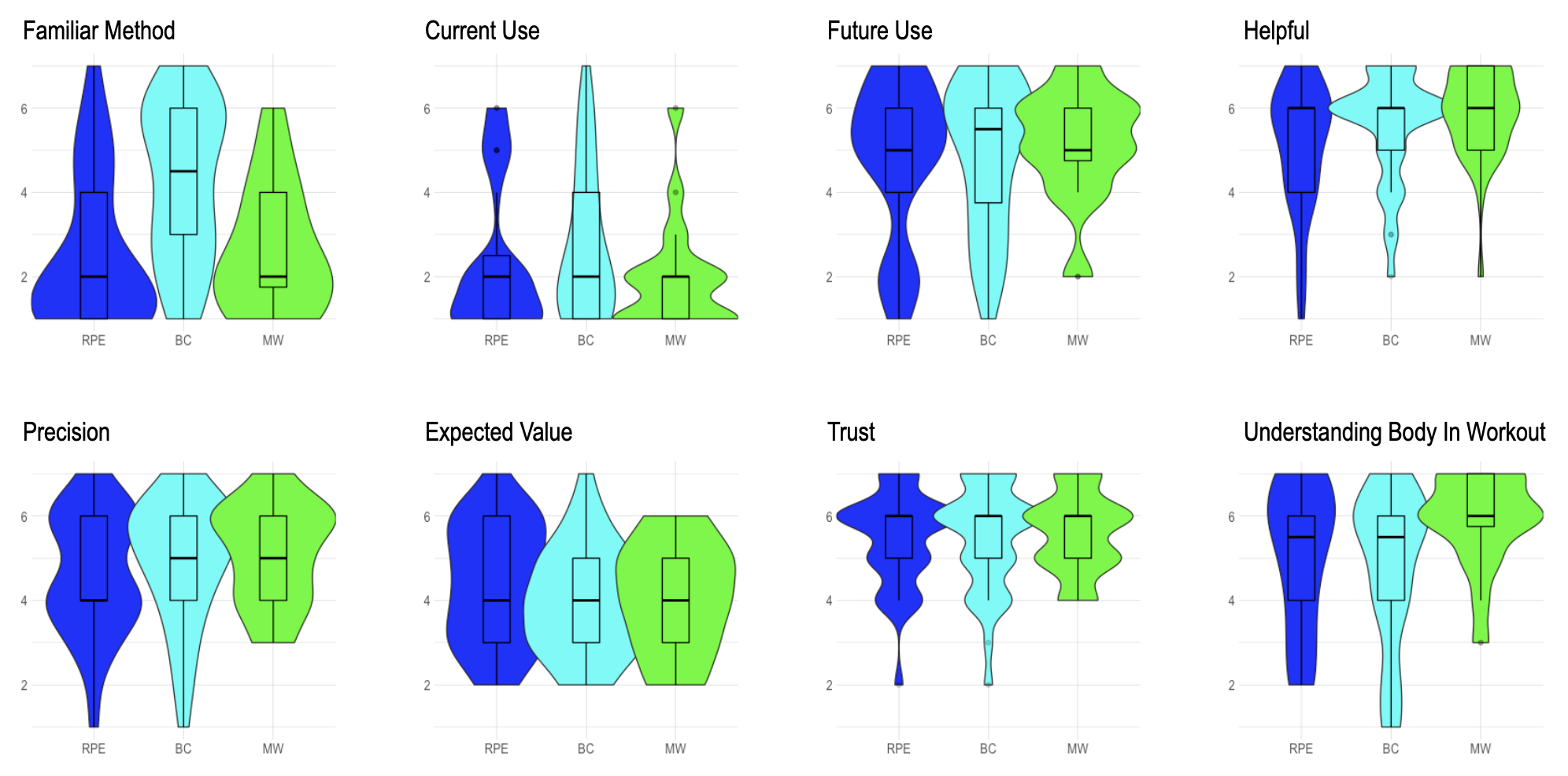}
    \caption{Plots depicting the results of 7-point Likert scale custom questions. The violin plots show the distribution of the respective variables. At the same time, a standard boxplot is drawn in the middle. }
    \Description{The figure shows eight violin-boxplots. Each plot shows the three conditions: RPE, burned calories and muscle work. In the first plot the "Familiar Method" is for burned calories very distributed with a bit higher than four, while RPE and muscle work are very low. For "current use" all values and means are low, only burned calories are stretched a little bit higher. For "future use" all measures are very high, mostly values between four and seven. For "Helpful" all measures are high as well, but you can see that RPE is a bit stretched down to lower levels while burned calories and muscle work have mostly high values starting from four. For "Precision" the plots look very similar with many distributions between four and six but interestingly, they start for muscle work at three and not below. For "expected value" the distribution is quite similar and mostly between two and six. For "Trust" the values are very high between four and seven with a mean around six but muscle work has no value lower than four. For "Understanding Body in Workout" the values are overall high, but for muscle work, there is an accumulation at six, while the other values are more stretched over the whole scale.}
    \label{fig:boxplotpart1}
\end{figure*}

\begin{figure}[!t]
    \centering
    \includegraphics[width=0.5\textwidth]{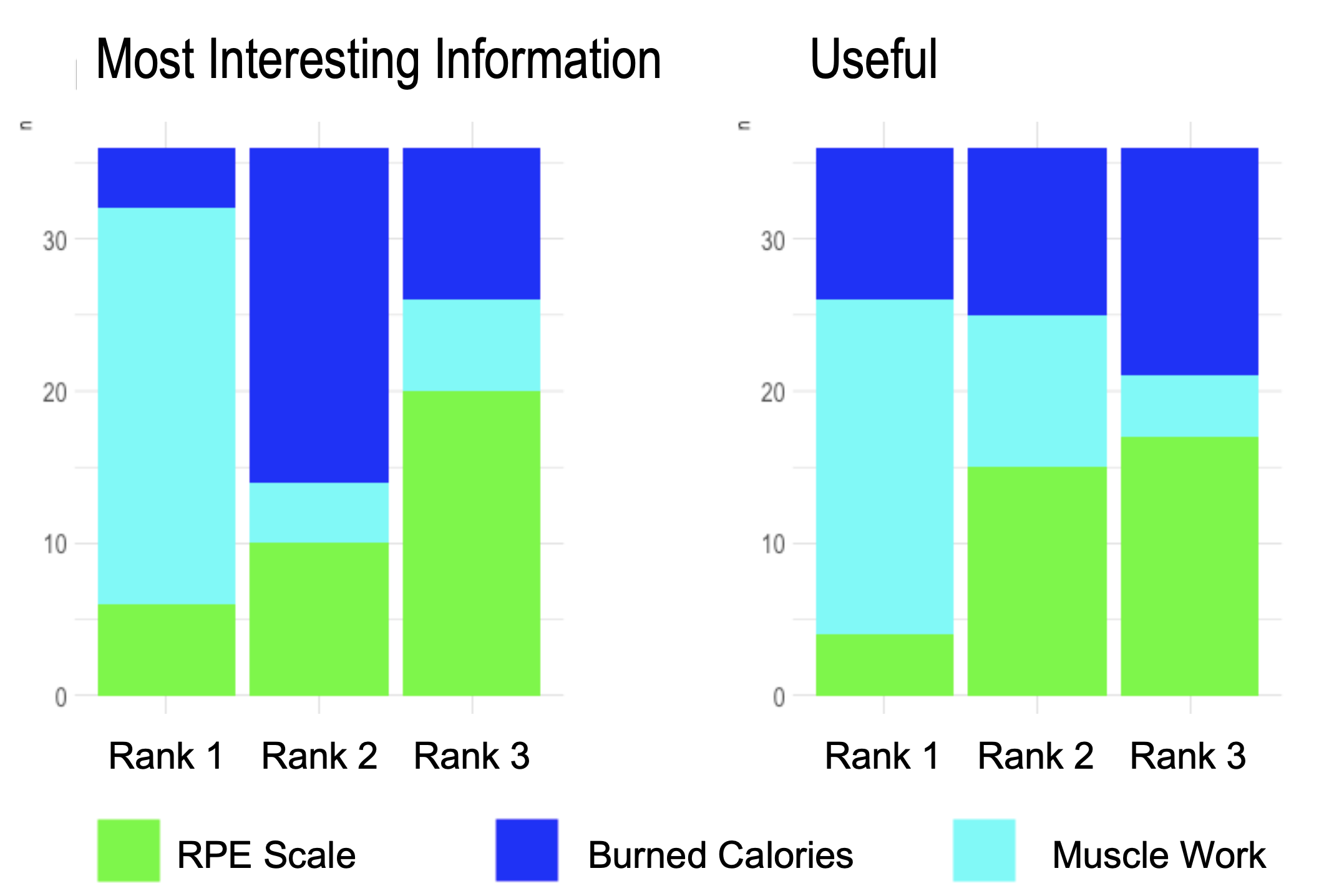}
    \caption{The user-chosen rankings of (1) RPE, (2) BC (3) MW with \EXMA, where the first rank is on the left and the third rank is on the right. We show two rankings, the left shows the measurement with the most interesting information, and the second one shows what kind of measurement is the more helpful one.}
    \Description{There are two bar charts showing three ranks on the x-axis for the distribution of decisions for the RPE scale, burned calories and muscle work. The first graph ranks the "most interesting information". It can be seen that the muscle work covers the most space in rank one, while the RPE scale covers a little above 50\% in rank three the lowest rank. The second scale is usefulness, where we can see the same pattern for muscle work high on rank one and RPE above 50\% for rank three. Burned calories, however, gained numbers in rank one than in the plot about the most interesting information.}
    \label{fig:rankings}
\end{figure}

\subsubsection{Procedure}

Participants were first introduced to the procedure and had to consent to their participation.
Afterwards, they were asked to put on the heart rate sensor and the Meta Quest 2 headset, so that everything fits comfortably. 
They first played the FitXR boxing tutorial. Afterwards, all values and log files were checked to see if everything was working well.
Then, the participants played one (the first) complete level in FitXR.  
After playing they put down the VR headset and used the keyboard for input on the large screen in front of them. 
In a counterbalanced order, they were presented with (1) the RPE Scale, where they had to choose their level, (2) the burned calories, which were calculated based on their weight and age input, or (3) the results of \EXMA~ (i.e., the muscle work) based on an avatar with highlighted muscle groups and bar charts. 
After each presentation of physical activity measurement, participants filled out the post-game questionnaire and finished after the three conditions with the rankings.
%When all three conditions were done, they answered some 

\subsection{Analysis and Results}

We statistically evaluated the eight custom single-item questions.
Since the distributions of all items are non-parametric, we use Friedmans Anova (fA) and report effect size with Kendall's w, mean (M) and standard deviation (SD). 
Values are rounded from two decimal places.
The only item showing statistical significance was the \textit{familiarity with method} item (fA = 16.8, p < 0.001, w = 0.234 small; RPE: M = 5.14, SD = 1.49 ; BC: M = 5.58, SD = 1.25; MW: M= 5.89, SD = 1.06).  
Post-hoc tests show significant differences between (2) Burned Calories $\leftrightarrow$ (1) RPE Scale (p.adj < 0.01) and  (2) Burned Calories $\leftrightarrow$ (3) Muscle Work (p.adj < 0.001). 

Non-significant results are \textit{current use or method} (fA = 4.4, p = 0.11; RPE: M =  2.36, SD = 1.76; BC: M = 2.75, SD = 1.78; MW: M = 1.97, SD = 1.3), \textit{future use of method} (fA = 1.35, p = 0.509; RPE: M = 4.80, SD = 1.85; BC: M = 4.89, SD = 1.75; MW: M = 5.14, SD = 1.31), \textit{helpfulness} (fA = 4.18, p = 0.124; RPE: M = 5.14, SD = 1.50; BC: M = 5.59, SD = 1.25; MW: M = 5.88, SD = 1.06), \textit{perceived precision} (fA = 3.5, p = 0.17; RPE: M = 4.64, SD = 1.44; BC: M = 4.94, SD = 1.47; MW: M = 5.11, SD = 1.19), \textit{expected value} (fA = 1.46, p = 0.48; RPE: M = 4.36, SD = 1.4; BC: M = 3.97, SD = 1.25; MW: M = 4.08, SD = 1.27), \textit{trust}(fA = 0.56, p = 0.76; RPE: M = 5.47, SD = 1.11; BC: M = 5.47, SD = 1.23; MW: M = 5.58, SD = 0.94), \textit{understaning body in workout} (fA = 4.84, p = 0.089; RPE: M = 5.06, SD = 1.67; BC: M = 4.83, SD = 1.82; MW: M = 5.88, SD = 1.09)

The associated violin-boxplots for all custom single items can be found in \autoref{fig:boxplotpart1}.\\

% Likert specific questions 
A set of questions was only asked for a specific measurement technique.
Participants rated the difficulty to use the RPE scale (on a 7-point-Likert scale) with M = 3.97, SD = 1.73.
For muscle work, they rated the understandability with M = 6.31, SD = 9.51, their wish for textual representation M = 2.53, SD = 1.32, and their wish for fixed value representation M = 4.16, SD = 1.54. 

We also checked the participants' understanding of the calculation of burned calories. 
Here, only one participant gave a reasonably correct answer to the question of how burned calories are calculated. %("the heart rate, age and body weight are variables to estimate the metabolism in a certain amount of time. So I guess there is a formula in which once set those values, then the total calories are calculated as the time passes by").
%(This question was asked after participants rated trust, etc. ) 
Only eight participants knew that heart rate and at least one other factor are involved in the calculation of burned calories. 
19 did not know and gave a wrong or inadequate answer, such as that energy or intensity of an exercise is measured. 

\begin{comment}
HR, Time, breathing, Intensity 
X- Energie / Intensity
W, HR 
W, Time, Metabolism 
X- Ausgeatmetes CO2 
W, HR
HR, Time 
X
X -Ration of mucle to fat - muscle activation to total fat 
X
X
X
W, HR, metabolic, Intensity
X
X -Energie / Intensity
X - Energie / Intensity 
HR, Blutdruck, Atemfrequenz
X
X - Amount Schwitzen
HR, speed of movement
X -> explains XERA
X
X
W, HR
W, HR
X- Intensity extrapolation
HR
X - Intensity
W, Time, Intensity
W, HR, AGE, Time , metabolism : 
W, HR
W, HR
W, HR, Intensity
HR
X - Intensity
X 
\end{comment}

% Combinations  
When we asked participants about the preferred combination of measures, seven chose burned calories with RPE, 16 chose the RPE scale with \EXMA, and 20 chose burned calories with \EXMA. 

Finally, we let participants rate the three measurements with regard to different factors. 
While most of the results are consistent with the results of the single-item questions, there are two interesting ratings, see \autoref{fig:rankings}.

\subsubsection{Qualitative Results}
We asked participants to comment on what they could think of to make each of the three methods more useful and which disadvantages and benefits of the method they saw. 
A sample of comments that seemed to be relevant to the paper authors are reported in this section. 
Some comments were translated from German into English with DeepL\footnote{\url{https://www.deepl.com/en/translator}}.

\paragraph{RPE}
Participants mentioned that there was ``too little information'' on the RPE scale and ``more precise information'' would be wished for. Participants also found that perception, in general, is very ``personal'' and ``you have to know your body to rate RPE''. 
On the other side, they thought that the RPE scale is a ``fast estimation'', helps ``to get a feeling of own exhaustion'', and that the ``observation of the training development [is] better possible''.
Someone explained that ``It is good to be able to compare your own perception of the challenge with the objective effort of the body''. 
To overcome the complexity participants would have liked ``examples'' and the possibility for ``comparisons to other participants''. 
Overall, participants saw the limitations of RPE: it only gives little or imprecise information, and it further needs a noticeable amount of training to actually yield realistic results and heighten the user's body awareness. 

\paragraph{Burned Calories}
Participants found that burned calories are ``easily comparable'', ``seem to be quite precise'', ``short'', and ``easy to use if you want to lose weight'' because one can ``calculate better what you have to eat''. However, someone pointed out, that the connection to food ``could lead to eating problems''. 
Simultaneously, some participants thought that burned calories are ``complicated'' and they ``might not be exact'' because ``heart rate could be increased by other factors''. 
Participants also criticized that it ``is hard to improve training performance if no explanation of the value is given'' and that burned calories ``give no information about what [one] could have done better''. 
Furthermore, participants stated that ``it is good for a cardio workout but not for training specific regions of the body''.
As improvements participants wished for ``more information'', ``an explanation of how it is calculated'', ``whether this value is good or bad'', and information about the ``way [I] can do the exercises more effectively''. 
%In summary, two groups of opinions manifested: users that valued the simplicity and potential ease of comparision, and users that saw challenges in precision and information density.
In summary, participants were divided between the benefits of comparability and simpleness towards imprecision and low information.

\paragraph{\EXMA}
When commenting on the \EXMA~ visualization, participants liked the ``clear representation'' which they found ``understandable for everybody'', that they could see ``which parts of [their] body were used heavily'' and that it allowed them to ``focus on [a] specific bodypart[s] while training''.
One person mentioned that a ``visualisation is better than only using the scientific names of muscle groups which none understands'' but at the same time ``people who are currently not so familiar with the body can learn which muscles are targeted and trained during certain exercises''.
One participant explained `` [...] you know how much a body part should move and where it is best to stretch before exercising to avoid injury.''. Another one further elaborated ``One can work on their weak spots, which means [one] can do better adjustments during the workout to train more those muscles that initially don´t get trained enough. It brings [the] opportunity to develop more muscles equally and even prevent some injuries (f.ex. overtraining of some muscle groups)''.
On the negative side, participants expressed reservations: ``I think that not my whole body was represented here'', and suspected that ``possibly the individual muscle groups cannot be distinguished exactly''.
They further mentioned that there were ``no clear values'' visible, which makes ``comparing two performances difficult''.
A participant mentioned that ``special equipment is needed'' likely referring to the camera and computer setup. 
One participant highlighted that ``this method is very focused on the body, which could be hard for people with more body fat''. 
For improvement, participants wished for a ``clarif[-ied] diagram with numbers'', ``multiple, more accurate levels in the scale/more colours'', representation of ``more muscle groups'', ``more interpretation'', and a ``comparison to other users''. 
Furthermore, they would have liked ``tips on how to improve'' and a ``comparison to a perfect example, how it should look like in contrast to how [they] did it''.
Altogether, participants saw several potential benefits of \EXMA, like preventing injuries, helping balance out a workout and learning about the body. At the same time, they pointed out the complexity of the results and setup, while giving valuable input for further improvements.  

\begin{figure}[!b]
    \centering
    \includegraphics[width=0.5\textwidth]{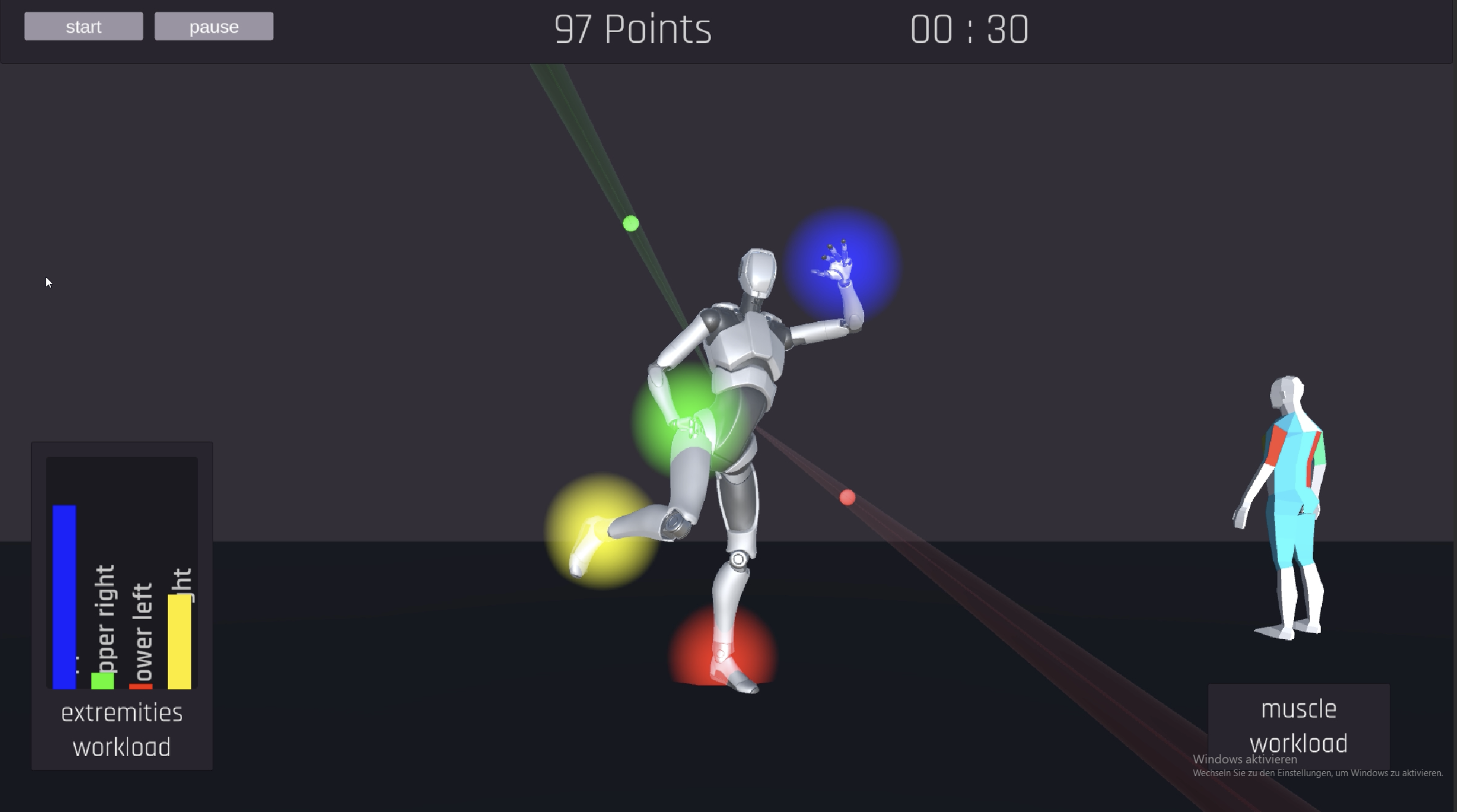}
    \caption{On the left and right of the screen, participants could see the visualization of \EXMA~ as a bar chart and 3D avatar. In the middle, an excerpt from the exergame can be seen. The avatar is mirroring the participant, who has to catch the balls flying around with the correspondingly coloured body part (e.g., red spheres have to be caught with the left leg, similarly highlighted in red).}
    \Description{The figure shows the visualization in the situation used for the second study. In the middle, there is an avatar in motion trying to catch spheres. The hand and feet are coloured in four colours (red, green, blue, yellow). The same coloured hand or feet has to catch the coloured sphere. The colours are also represented at the left in a bar chart showing the amount of muscle work for each limb. It can be seen that the left arm (blue) and right foot (yellow) are lifted, so the bar chart for blue and yellow is high as well. On the right, there is an avatar where the left upper arm and the side of the body are highlighted red since these must be the body parts most involved in this movement. Probably there are other body parts highlighted in red, but this can not be seen from this perspective, since the avatar is normally turning to show all perspectives.}
    \label{fig:SphereExergame}
\end{figure}

\subsection{Measure Comparison}
%Comparison of the three measured values

\begin{figure*}[h]
    \centering
    \includegraphics[width=1.00\textwidth]{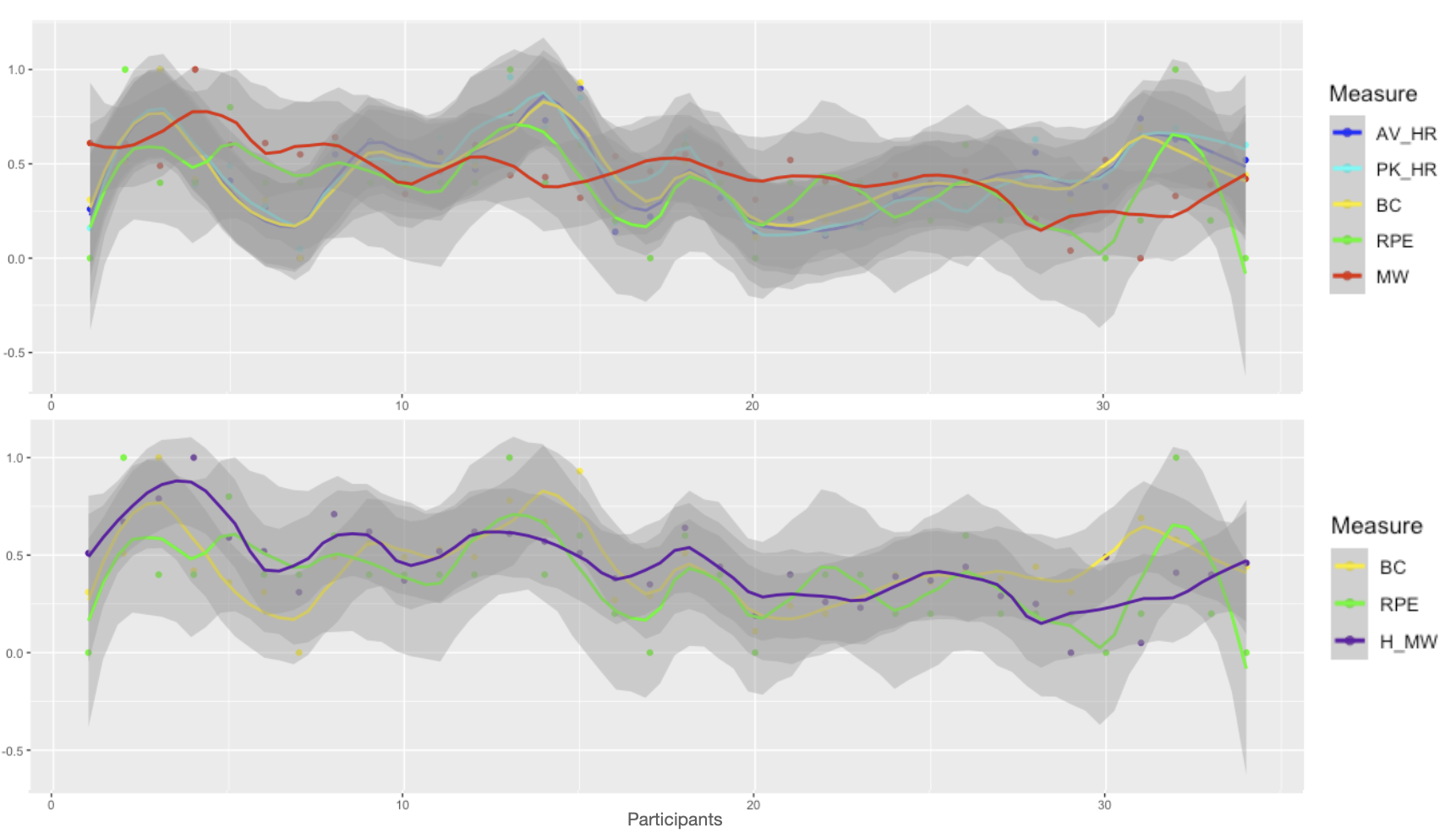}
    \caption{The upper graph shows normalized values (y-axis: 0 to 1) of average heart rate (AV\_HR), peak heart rate (PK\_HR), burned calories (BC), RPE scale, and muscle work (MW). for each participant (x-axis). The lower graph shows the same normalization for participants but for burned calories (BC), RPE scale, and a muscle work value that includes heart rate as a factor (H\_MW). We can see, that the H\_MW develops closer to burned calories now, but still, all three measures develop slightly differently.}
    \Description{The figure shows two scatter plots where on the x-axis we have each participant and on the y-axis, there is the normalized value between 0 and 1 for in plot one: average heart rate, peak heart rate, burned calories, RPE and muscle work. The values are slightly smoothed out and show a graph for all four values. The graph of average heart rate, peak heart rate and burned calories are almost similar, while the other two are different. For some values, some curves match but then for other participants, they do not match again. 
    For plot two there are values for burned calories and RPE and muscle work with heart rate as a factor. Here we have smoothed line graphs as well. The three lines differ but we can see that including heart rate as a factor in muscle work approaches the burned calories curve.}
    \label{fig:ExertionMeasures}
\end{figure*}

Besides our observations that \EXMA~ values change according to movement executions look fitting, we were interested in how the different exhaustion measurement techniques developed for each participant individually. 
In \autoref{fig:ExertionMeasures} we visualize the normalized values of average heart rate ($AV\_HR$), peak heart rate ($PK\_HR$), burned calories ($BC$), RPE scale ($RPE$), and muscle work ($MW$) for each participant while playing FitXR. 
We can see that the two heart rate values and burned calories develop very similarly, which makes sense since burned calories also include heart rate as a factor.
Self-perception with the RPE scale develops differently as well as muscle work.
In the second plot, we again report $BC$ and $RPE$ normalized values for each participant but add heart rate as a factor to muscle work and then normalize it.  
As expected, the curve approximates the burned calories curve, which includes the heart rate as well. 

Thus, it is not possible to say whether when participants perceived effort to be low, burned calories and muscle work were low, too, as opposed to participants who perceived effort to be high and also measured high burned calories and muscle work values.
It seems, therefore, not comparable whether a measurement technique develops similarly to another one and thus could indicate more suitable results.

\section{Discussion \& Future Work} % 1,5 Spalten Doppelseitiges Format 

We could show the feasibility of our system to measure the muscle work of different muscles and visualize them with a suitable level of information: it is perceived as helpful, with a level of granularity that is detailed yet not complex.
While this demonstrated the applicability of our approach, it has to be noted that the Kinesis muscle model is a generic representation of a human being without the option for individual manifestations. This is particularly important when considering that women have a different body structure than the male body the system is based on. 
However, in our opinion, we have obtained very convincing data for the purpose of exergames. 
For other use cases such as dancing, the feasibility has to be evaluated in more detail with respective experts in this sport. 

\paragraph{Technical Validity}
Our observations and technical evaluation confirmed that the exertion values for a muscle correlated with the muscles that participate in the corresponding movement. Whether the values are relative to each other and represent fitting height differences can not be definitively confirmed as the calculations are based on our model, which omits small details that are people independent (e.g., slightly different muscle lengths). 
Nevertheless, our technical validation shows that the values are more or less correct. In confirmation with our physiotherapist, the values are fitting with respect to the exercises, apart from the overall higher upper limb to lower limb values. Here, however, we propose to improve the tracking method since we are convinced that the values come from the tracking glitches. 
Furthermore, we demonstrate in our technical evaluation that a person performing two different measures of squats (with and without arms) induces almost the same values for each muscle, except of the arm related muscles. The small difference occurs due to natural fluctuations due to tracking and execution differences.
In future work, individual aspects of the person (age, weight, heart rate) should be included and the model and calculation should be adapted.
For example, this would include that muscles have to work ``harder''(i.e., \emph{more}) if the user's body weight is higher. 
Possibly, even further individual characteristics can be included with a detailed body scan.

\paragraph{Comparison to RPE Scale and Burned Calories}
%We could see through observations that the values of muscles involved in a movement also increased. 
Since it is not possible to compare the values (burned calories, RPE, muscle-work) on the same scale, we normalized them to understand if they correlate (i.e., if high heart rate values also lead to high RPE and \EXMA~ values).
As we did not observe a close relation between any of the three values, we can conclude that they each provide complementary information that is important for a person to understand their performance.

%What we could do, however, is to compare the different measured values in the same scale to see if high heart rate values also led to high RPE or \EXMA~ values.
Our results in~\autoref{fig:ExertionMeasures} show only a connection between heart rate values and burned calories, since the calculation of burned calories is based on heart rate values. 
%In the second graph, we show that applying heart rate to \EXMA~ causes the curve to converge to the burned calories curve. 
%In future work, we could think about a formula similar to burned calories to better include individual differences of participants.
For future work, we consider it promising to integrate burned calories and muscle work into one measure. We depicted a version of that in the second graph, where we applied the heart rate value to \EXMA~.
Nevertheless, heart rate as a factor should be considered with caution, since this value can also be influenced by other causes (e.g., excitement or fear, applicable particularly in exergame contexts).

Currently, it is not possible to say which value is more accurate, but only that they sometimes differ greatly. The reason for this must be further evaluated.

%Since we could not find clear converged curves between RPE Scale, burned calories, and muscle work in \autoref{fig:ExertionMeasures}, we can not claim that values fro mall the measurement techniques emerge in a similar expression. 
%However, we can say from both plots in \autoref{fig:ExertionMeasures}, that heart rate as a factor strongly influences objective exertion measures. 

%* technisch schwer zu sagen ob genauer wie BC oder RPE aber soll ja auch nicht ersetzten, participants mögen ein komination und daher kann es einfach auch für granularere informationene genutzt werden 
\paragraph{Participants' perception between RPE, Burned Calories and Muscle Work}
Since burned calories are an established method used in many fitness watches and devices, it is not surprising that burned calories were rated to be significantly more familiar to participants. 
However, it is interesting that participants indicated trusting all three methods even if they did not understand how they are calculated (e.g., in the case of burned calories).
Participants indicated that they would like to use all three methods in the future.
Participants also noted that all methods seemed precise and familiar, and they helped participants understand their workouts. 
Overall, we can say that all of the methods have their justification, but when participants had to choose, they decided that \EXMA~ had the most interesting information and was therefore rated as the most useful method. 
 \EXMA's benefits are further detailed in the next paragraph. 

%* burned calories is known but participants have mostly no idea how it works -> our tool no lack in perceived trust, precision, helpfulness 
%* most interesting and useful information has our tool 

%As it is hard to say what measurement technique is more accurate, we can only say that \EXMA includes the change of motion each second, while burned calories and RPE use approximate values. 
%Still, we do not want to argue which method could be better, we rather assume that the methods can complement each other like it was also suggested by participants.  

%* technisch schwer zu sagen ob genauer wie BC oder RPE aber soll ja auch nicht ersetzten, participants mögen ein komination und daher kann es einfach auch für granularere informationene genutzt werden 

\paragraph{Benefits of \EXMA}
\EXMA~ is seen as very useful and supportive to understanding the body during or after a workout (in particular, the exertion and understanding of the body parts and which parts moved most). 
Participants would use \EXMA~ in the future, think of it as being precise and trustworthy and it behaves somewhat as expected.  
The NASA-TLX score and game performance score suggest that the in-situ visualization is helpful and does not distract participants from their exercise task.
However, future work needs to explore to what extent the visualization should be displayed constantly, only for specific occurrences, or exclusively ex-situ for monitoring and analysis purposes.
%Whether it really makes sense to constantly display the visualization and changes during the workout still needs to be evaluated in more detail. 
Participants also see the advantage of displaying the muscle work of different muscle groups to prevent injuries and excessive strain and support a balanced workout while staying motivated for the next workout.
For future work, it could be very promising to adapt this measurement to regularly monitor any kind of stationary exercise.

%* \EXMA visualization not distracting, useful (auch statistisch) for understanding body in workout and balanced workout -> health , motivation, 

Because the muscle work measurement is camera-based and includes additional hardware, it only works for stationary setups but it can be easily transferred to different places since it is small and not very heavy. 
To have even more mobility and lower extra costs, the use of smartphone cameras would improve our approach further. The movement tracking with the Azure Kinect depth cameras was very promising.
However, current developments of smartphone-based movement tracking open up the possibility to transfer our system to smartphone-based solutions in future work. 
%* funktioniert nur stationär nicht für cardio/ joggen usw. 
%* additional teure hardware aber einfach änderbar

\paragraph{\EXMA~ Visualization}
Most participants liked both representations (3D avatar and bar chart). They found it easy to understand even for inexperienced users and believed it could help to learn and understand muscles and the body. 
Nevertheless, there is room for improvement in future work. 
The muscle groups can be visualized with more details, the detail of which could be chosen by the users. % and left to users to choose the degree of detail themselves. 
In addition, the participants were interested in comparability, for which exact values for each muscle, muscle group, and body part need to be displayed. 
A complete application that stores values over time and allows further extensions is planned as future work. 
%-> zukunft technisch verbesserung und userfull visualizations erweitern 

\subsection{Limitations}

Regarding the technical limitations, we are aware that our system is not 100 percent precise. 
It has some weaknesses, like the camera-based tracking that differs between clothing colours or camera angles. The implementation itself similarly has room for improvement, where we do not take height, BMI, individual muscle length, and bone structure into account. 
However, we were able to show that XEM generally works and could be suitable for personal training monitoring, where it can be used as an approximate benchmark and provides more detailed information than RPE or BC.

%Furthermore, the measurement with a camera is not perfect, there are still jitter effects that must be compensated for in the data and that make the measurement less precise. 
%Moreover, the Azure Kinect cameras have problems detecting the body if participants wear something with poor contrast with the background. 
%Besides, we currently cannot provide exact real-time results (only with small offset), but since participants were not overly interested in an in-situ solution anyway, a post-training display might be sufficient.

As the study was conducted with exergames as a use case, we cannot directly apply the statements to ``normal'' home workouts or dancing sessions, but only assume that they apply similarly. Furthermore, participants were recruited in the university environment and probably have an interest in sports and technology-related topics, which could have influenced the data.

VR novelty or novelty of the measurement techniques also likely influenced participants' enjoyment and usage of \EXMA. 
Novelty effects could have influenced the exposure either positively (novel systems can be interesting and enjoyable \cite{rodriguesGamificationSuffersNovelty2022a,ruttenInitialEncounterMidAir2021}) or negatively (users being suspicious about new techniques \cite{alrajaEffectSecurityPrivacy2019a}). However, these effects can exist in any study where participants are exposed to novel systems. With our additional qualitative findings, we see that participants indeed see a benefit in the new technique, so we remain confident that \EXMA~'s value was appreciated beyond its novelty.
Furthermore, our results suggest no influence of the values from the different measurements since there was only one exposure in VR before answering questions regarding all three measurement techniques.

The code for the technical validation is slightly different than from the usability study, since we found a minor implementation error.
Before, some muscles that operate in a different direction induced negative values, which were subtracted instead of added. The difference in the results was not major but visible, especially in the shoulder part.

\section{Conclusion}
In this paper, we presented a novel approach to measure exertion based on muscle work for stationary physical exercises with a motion-tracking setup that remains reasonably simple and transportable. 
We presented \EXMA, a camera-based exertion monitoring approach to calculate muscle work of different muscle groups and visualize them through bar charts or using a 3D avatar model with highlighted muscle groups. 
In our study, we were able to show that measuring muscle work with the Azure Kinect is feasible and that monitoring exercises for each muscle group is an information gain that users see as helpful, motivating, and supportive. %would support their exercise. 
For future research, expanding the user interface with more granular and time-dependent information and enhancing the measurement accuracy through better tracking would be favourable.
Ultimately, approaches like \EXMA~ may strike a balance between feedback simplicity and usefulness while presenting the feedback during the exercise instead of after.

%%
%% The acknowledgments section is defined using the "acks" environment
%% (and NOT an unnumbered section). This ensures the proper
%% identification of the section in the article metadata, and the
%% consistent spelling of the heading.
\begin{acks}

\end{acks}

%%
%% The next two lines define the bibliography style to be used, and
%% the bibliography file.
\bibliographystyle{ACM-Reference-Format}
\bibliography{MuscleWork}

%%
%% If your work has an appendix, this is the place to put it.
%\appendix

\end{document}